\documentclass[preprint,aps,nofootinbib,amsmath,amssymb]{revtex4}
\usepackage{latexsym}
\usepackage[dvipdfm]{graphicx}
\begin{document}
\title{ Radiatively Induced Spontaneous Symmetry Breaking by Wilson Line
in a Warped Extra Dimension}
\author{Hisaki Hatanaka}
%
%
\affiliation{Department of Physics, Chung-Yuan Christian University,
Chungli, Taiwan 320, R.O.C.}
\date{December 9, 2007}
%
%
%
\begin{abstract}
We investigate the dynamical gauge-Higgs unification in the
Randall-Sundrum (RS) space-time. We study the dynamical gauge-Higgs
unification in the $SU(2)$ gauge theory with a bulk fermion in the RS
space-time.  We evaluate the contribution from fermion loop to the
one-loop effective potential with respect to the Wilson-line phase, and
study the dynamical gauge symmetry breaking. We also apply this
mechanism of the gauge symmetry breaking to the electroweak gauge-Higgs
unification in the RS space-time. Especially we numerically studied a
$SU(3)_w$ gauge model as a toy model of electroweak gauge-Higgs
unification in the RS space-time. We introduce an adjoint fermion into
the model to break the gauge symmetry and to obtain the $U(1)_{\rm em}$
electromagnetic symmetry.  We found that in this model the ratio of
$Z$-boson mass to $W$-boson varies with respect to the Wilson-line phase
even at the tree level.  We also propose a dynamical mechanism of tuning
the ratio $m_Z/m_W$ to the experimental value
$91.2\text{GeV}/80.4\text{GeV}=1.13$ by introducing bulk scalars or bulk
fermions with twisted boundary conditions.  In these models the Higgs
can vary in mass between zero and $290\,\text{GeV}$.
\end{abstract}
\maketitle
%
%
\newcommand{\tr}{\mathrm{tr}}
\newcommand{\diag}{\mathrm{diag}}
\newcommand{\thetaH}{\theta_H}
\renewcommand{\Re}{\mathrm{Re}}
\renewcommand{\L}{{\cal L}}
\newcommand{\Li}{\mathrm{Li}}
\newcommand{\hatm}{\hat{m}}
\newcommand{\C}{{\cal C}}
\newcommand{\ad}{\mathrm{ad}}
\newcommand{\fd}{\mathrm{fd}}
\newcommand{\gh}{\mathrm{gh}}
\newcommand{\eff}{\mathrm{eff}}
\newcommand{\mubar}{\bar{\mu}}
\newcommand{\GeV}{\text{GeV}}
\newcommand{\half}{{\textstyle\frac{1}{2}}}
%
%
\section{Introduction}
%
%
The standard model has been in good agreement with current experimental
data.
The origin of the electroweak symmetry breaking, however, has not yet
confirmed.
As an alternative to the Higgs mechanism, the electroweak gauge-Higgs
unification (EWGHU) has been considered for many years (for early works,
see \cite{Fairlie,Manton,Krasnikov}).
In the EWGHU, the Higgs field is regarded as an extra-dimensional
component of the gauge field in higher-dimensional space-time.
If the gauge group is non-Abelian and the spatial extra dimension is
multiply-connected, then the gauge symmetry can be broken by the
non-Abelian Wilson line phase in the extra dimension (i.e., broken
through the Hosotani mechanism) \cite{Hosotanimechanism,Witten85}.
Furthermore, such Wilson line phases can be dynamically induced
\cite{Hosotanimechanism,Hosotani89,Hosotani02}. One can define the
effective potential with respect to the Wilson-line phases. The
Wilson-line phases are dynamically selected as the phases that minimize
the effective potential.
This dynamical mechanism of the gauge-symmetry breaking by
Wilson-line phases is referred as the dynamical gauge-Higgs unification.
In the dynamical gauge-Higgs unification, both of the size of the mass
and the magnitudes of the effective potential are almost of the order of
size of the extra dimensions.
Therefore, since possibilities of the TeV-scale extra dimensions are
pointed out \cite{Antoniadis90,LED}, the EWGHU scenario has been
extensively studied in the context of the dynamical gauge-Higgs
unification
\cite{HIL,Nomura,Antoniadis_Benakli,Choi,Panico,Antoniadis-Benakli-Quiros,Lim-Kubo-Yamashita,Hosotani04,bulkmass,Murayama,Wudka06,finiteness,Haba-Takenaga-Yamashita,Sakamoto06,Cacciapaglia,GHU-Yukawa,EWGHU_others,adjoint-fermion}.
%

%
However, in the construction of a realistic model of the EWGHU in the
flat extra dimension, we encounter several obstacles.
To see this, let us consider the models with ``flat extra dimensions''.
Here models with flat extra dimensions are models such that
(i) Extra dimensional space has vanishing curvature (e.g., a circle
$S^1$ or $n$-torus $T^n$).
(ii) The metric of the spacetime is assumed to be ``factorizable'', in
other words, the metric of the four-dimensional space-time can be
independent of coordinates of extra dimensions.
The obstacles we will see in these models are as follows.
\begin{itemize}
\item[(a)]\label{light-Higgs} In models of the EWGHU with dynamical
	  gauge-Higgs unification in a flat extra dimension, we obtain
	  too small Higgs mass.
In \cite{Hosotani04} the Higgs mass $m_h$ is estimated to be $m_h = {\cal
O}(m_W \sqrt{\alpha_W})$, where $\alpha_W \equiv g_4^2/4\pi$ with the
four-dimensional (4D) weak coupling $g_4$.
With $m_W=80.4\,\GeV$ and $\alpha_W=0.032$, we obtain $m_h ={\cal
	  O}(10)\,\GeV$, which contradicts the
	  observation.\footnote{Ways of pushing up the Higgs mass by
	  fine-tuning of the Wilson-line phase are studied in
	  \cite{Haba-Takenaga-Yamashita}.}

\item[(b)]\label{universal-Yukawa} In models of the EWGHU in the flat
	  extra dimension, the mass of the quarks and leptons tend to be
	  universal and the mixing matrix of quarks to be diagonal,
	  because the gauge coupling constant yields the ``Yukawa
	  couplings'' of fermions to the Higgs in the gauge-Higgs
	  unification.
          Therefore, it needs some other mechanisms to induce large mass
	  hierarchy and mixings among fermions.\footnote{Some ways to
	  obtain the Yukawa hierarchy in the gauge-Higgs unification
	  are discussed in \cite{Murayama,GHU-Yukawa}.}

\item[(c)]\label{light-flavor} In the models in the flat extra dimension,
	  one of the simplest way to give different masses to fermions is
	  to assign the different bulk mass term to each fermion.
Then, in the EWGHU the mass of the lowest Kaluza-Klein (KK) mode of such
	  a bulk fermion is estimated to be $\sqrt{\theta^2/R^2 +
	  m_f^2}$, with the Wilson line phase $\theta$, compactification
	  radius $R$ and bulk mass of the fermion $m_f$.
Thus, a fermion with larger (smaller) bulk mass can be heavier (lighter)
	  in the four dimensional effective theory.
Since the effective potential decrease in size with increasing the bulk mass
  of the fermion \cite{HIL}, it lead us to an odd conclusion:
 {\it a light fermion
  (e.g., up-,down-quark or electron) has large contribution to
  the effective potential of the Higgs.}
The low energy dynamics of the EWGHU in the flat extra dimension looks
	  very different from 4D models of the electroweak symmetry
	  breaking by the Coleman-Weinberg (CW) mechanism \cite{CW}.
\end{itemize}

%
In recent years, models of electroweak gauge-Higgs unification in the
Randall-Sundrum (RS) space-time \cite{RS} have been extensively studied
\cite{Toms,Oda,Medina,Hosotani05,Hosotani06,Sakamura06,Contino03,AdS-deconstruct}.
We first summarize the merits of the dynamical gauge-Higgs unification in
the RS space-time. They are
\begin{itemize}

\item[(A)] The hierarchy between electroweak and a larger scale (e.g.,
	   the Planck scale) are explained by the nature of the RS
	   space-time
      \cite{RS}\footnote{Attempts to obtain the large hierarchy within
	   the
      framework of the flat extra dimension are seen in
      \cite{Haba-Takenaga-Yamashita,Sakamoto06}.}, whereas the quadratic
	   divergence of the Higgs mass is absent due to the
	   higher-dimensional gauge symmetry
      \cite{Hosotanimechanism,HIL,finiteness}.

\item[(B)] The extra dimension of the RS has the orbifold topology
	   $S^1/Z_2$. On such space-time, we can naturally obtain
      the chiral structure of fermions by boundary conditions at the
	   fixed points. $Z_2$ projection also yields the Higgs field in
	   the fundamental
      representation of $SU(2)_w$ from the gauge field in the adjoint
      representation of a gauge group which includes the $SU(2)_w \times
	   U(1)$.

\item[(C)] In the RS space-time, we can obtain large mass differences
	   among fermions by tuning bulk mass parameters of fermions of
	   order of unity \cite{Neubert}.
In the same way we can obtain large mass hierarchy among fermions in
	   the EWGHU in the RS space-time \cite{Hosotani06}.

\item[(D)] The mass of the Higgs will be lifted up to
     a few hundred \GeV{} when we consider the dynamical gauge-Higgs
	   unification in the RS space-time \cite{Hosotani05}.

\item[(E)] When we implement the standard model in the RS space-time,
     the Higgs field (i.e., the 5th-dimensional component of the gauge
	   field) is naturally localized on the TeV brane. We note that
	   such a localization of the Higgs on the TeV brane is usually
	   required in the context of the 5D extension of the standard
	   model in the RS space-time
	   \cite{GP,RSbulk,Higgs-Localization}.
\end{itemize}
%
\par
The dynamical gauge-Higgs unification in the Randall-Sundrum space-time
is first considered in \cite{Toms} and effective potential by fermion
without bulk mass is evaluated in it.
An appropriate form of the background gauge in the RS space-time and the
calculation of the effective potential from gauge-ghost loop is studied
in \cite{Oda}.
The phenomenological studies of the EWGHU in the RS space-time are seen
in \cite{Hosotani05,Hosotani06,Sakamura06,Medina}.
We note that there are series of works \cite{Contino03}, based on the
AdS/CFT dual or ``holographic'' picture \cite{AdSCFT}.
The Wilson-line dynamics in the warped extra dimension has also been
studied in \cite{AdS-deconstruct} by means of the dimensional
deconstruction \cite{deconstruction}.
%
\par
%
In the present paper we investigate the dynamical gauge-Higgs
unification in the Randall-Sundrum space-time. The aim of the present
paper is the following threefold.
First, we evaluate the effective potential of Wilson line and discuss
the gauge symmetry breaking by the dynamical gauge-Higgs unification and
to evaluate the mass of the Higgs field.
Especially we are interested in the one-loop effective potential for 
the loop of fermion with a bulk mass term.
Second, we make clear the relation between the fermion mass
spectrum and the effective potential in this model.
In the case of warped extra dimension, the lowest KK mass of the fermion
decreases in size with increasing absolute value of the bulk mass.
Therefore we expect that in the RS space-time, a heavy fermion in the 4D
effective theory will have large contribution to the effective potential
and a light one has small contribution, as we have seen in the CW
mechanism. We will demonstrate it is true.
Third, we use the dynamical gauge-Higgs unification in the RS space-time
to construct a model of the EWGHU with realistic mass spectrum of
fermions and a Higgs potential.
%
\par
The present paper is organized as follows.
In Section \ref{sec-hosotanimech},
we study the $SU(2)$ gauge model in the RS space-time and evaluate the
one-loop effective potentials from the fermion loop and see how the
effective potential and the gauge symmetry depends on the bulk muss term
of the fermion.
In Section \ref{sec-su3model}, 
as an application of the dynamical gauge-Higgs unification in the RS
space-time, we reconsider the $SU(3)_w$ gauge model as a toy model of
EWGHU in the warped extra dimension.
Section \ref{sec-summary} is devoted to the summary and comments.
In Appendix \ref{sec-approx}  Approximation formulas
of low-energy mass spectrum of fermions are collected.
%
\section{Dynamical Gauge-Higgs Unification in the RS Space-Time}
\label{sec-hosotanimech}
%
\par
We consider the gauge theory in the the RS two-brane model \cite{RS}.
The RS space-time is a slice of five-dimensional anti-de Sitter
space-time $AdS_5$, and the metric of the RS space-time can be written
as
\begin{eqnarray}
ds^2 = G_{MN} dX^M dX^N 
=
e^{-2\sigma(y)} \eta_{\mu\nu}dx^\mu dx^\nu - dy^2,
\label{RS}
\end{eqnarray}
where the five-dimensional coordinates are $X^M = (X^\mu = x^\mu,
X^4\equiv y)$ ($M,N=0,1,2,3,4$ and $\mu,\nu=0,1,2,3$).
$\eta_{\mu\nu}$ is the 4-dimensional metric and $\eta_{\mu\nu} =
\diag(+1,-1,-1,-1)$.
We assume that the function $\sigma(y)$ has the periodicity
$\sigma(y+2\pi R) = \sigma(y)$ and the reflection symmetry at
$\sigma(y_i+y) = \sigma(y_i-y)$ ($i=0,1$) with $ y = y_0 \equiv 0$ and
$y = y_1 \equiv \pi R$, where $R$ is the compactification radius.
In the present paper we assume that $R$ is already fixed by some
mechanisms (e.g. Goldberger-Wise mechanism \cite{GW}).
For $-\pi R \le y \le \pi R$, we write $\sigma(y)=k|y|$, where $k$ is
referred as the curvature of the $AdS_5$.
Two boundaries $y = y_0$ and $y = y_1$ are referred as ``Planck
brane'' and ``TeV brane'', respectively.
\par
We consider a $SU(2)$ gauge theory in this space-time.
We also include a Dirac fermion $\Psi$ in the fundamental
representation, or $\Lambda$ in the adjoint representation.
The action of the gauge field and the fermion in the bulk space-time can
be given by $S_{\mathrm{gauge}} + S_{\Psi(\Lambda)}$, where
$S_{\mathrm{gauge}}$ and $S_\psi$, ($\psi=\Psi,\Lambda$) are given by
\begin{eqnarray}
S_{\mathrm{gauge}} &=&
-\int d^4x \int_{-\pi R}^{\pi R} dy \sqrt{|G|} 
\left[
 \frac{1}{4}g^{MN}g^{RS}F_{MR}F_{NS}
 + \L_{\mathrm{gf}} + \L_{\mathrm{ghost}}
\right],
\\
S_{\psi} &=& 
\int d^4 x \int_{-\pi R}^{\pi R} dy |E|
 \left\{
  i \bar{\psi} E^M{}_m \gamma^m D_M \Psi 
- k c_\psi \epsilon(y) \bar{\psi}\psi
\right\},
\end{eqnarray}
where $F_{MN}$ is the field strength :$F_{MN}=\partial_M A_M -
\partial_N A_M - ig[A_M,A_N]$ and $A_M$ is the gauge field.
$\L_{\mathrm{gf}}$ and $\L_{\mathrm{ghost}}$ are the gauge-fixing and
the Faddeev-Popov Lagrangian, respectively.
The five dimensional $\gamma$-matrices $\gamma^m$ are defined by 
$\left\{ \gamma_m, \gamma_n \right\} = 2\diag(+1,-1,-1,-1,-1) $ (Small
Latin indices $m,n$ represent Lorentz indices and $m,n=0,1,2,3,4$), and
$\gamma^{4} = -\gamma_4 = -i\gamma_5$.
$E^M{}_m$ is an inverse of the f\"unfbein.
$|G|$ and $|E|$ are absolute values of determinants of the metric tensor
and the f\"unfbein, respectively and $|E| = \sqrt{|G|}$ is satisfied.
$c_{\Psi}$ and $c_{\Lambda}$ are dimensionless bulk mass parameters for
$\Psi$ and $\Lambda$, respectively.
$\epsilon(y)$ is the sign function: $\epsilon(y) = +1(-1)$ for $y>0$
($y<0$).
We decompose a field $\Phi_{\ad}$ in the adjoint representation and
$\Phi_{\fd}$ in the fundamental representation into
$\Phi_{\ad} = \sum_{a=1}^{3} \sigma_i \Phi_{\ad}^{(a)}/2$
and
$\Phi_{\fd} = ( \Phi_{\fd}^{(i=1)} , \Phi_{\fd}^{(i=2)} )^T$
($\sigma_i$ is the Pauli matrices), respectively.
\par
As a gauge fixing, we choose the background gauge.
We separate the gauge field into background $A_M^c$ and quantum
fluctuation $A_M^q$: $ A_M = A_M^c + A_M^q $.
In the background gauge we write covariant derivatives of the gauge
field, fermions in the fundamental and adjoint representation as
\begin{eqnarray}
D^c_M A^q_N 
&=& \partial_M A^q_N - ig \left[A_M^c, A^q_N \right],
\\
D^c_M \Lambda 
&=& \partial_M 
    + \frac{1}{8}\omega_M^{mn} \left[\gamma_m,\gamma_n\right]
    -ig \left[A^c_M, \Lambda \right],
\\
D^c_M \Psi
&=& \partial_M
    + \frac{1}{8}\omega_M^{mn}\left[\gamma_m,\gamma_n\right] -ig A^c_M \Psi,
\end{eqnarray}
where $\omega_M$ is the spin connection and $\omega_\mu^{n4} = -
\omega^{4n}_{\mu}= -\sigma' e^{-\sigma} \delta^n_\mu$ ($n=0,1,2,3$), and
other components vanish.
We expect only $y$-component of the gauge field can have non-vanishing vacuum
expectation value(VEV), i.e., $A^c_M = \delta_{M}^y \langle A_y \rangle$.
By choosing appropriate gauge fixing term \cite{Oda} and corresponding
ghost term in the action, we obtain the quadratic action of $A_M^q$ and
the ghost field $\omega$ :
\begin{eqnarray}
S_{\mathrm{g}} 
&=& -
\int \!\! d^4x \!\! \int \!\! dy \, \tr
\left[
  A^{q\mu}(\Box - {\cal D}^2)A_{\mu}^q
+ e^{-2\sigma}
 A_y^q (\Box - D_y^c D_y^c e^{-2\sigma})A_y^q
+ e^{-2\sigma}
 \bar{\omega}(\Box - {\cal D}^2)\omega
\right],
\end{eqnarray}
where $\Box \equiv \eta^{\bar{m}\bar{n}} \partial_{\bar{m}} \partial_{\bar{n}}$ ($\bar{m},\bar{n}=0,1,2,3$)
and ${\cal D}^2 \equiv D_y^c e^{-2\sigma} D_y^c$,
respectively.
For later use, we introduce a new coordinate $z$, which is defined by
$z \equiv e^{\sigma(y)}$ ($0 \le y \le \pi R$).
With this coordinate, equations of motion for gauge fields $A^q_\mu$,
$A^q_z$ ($A_z = A_y/kz$)
and fermionic fields $\psi_{L,R} = \Lambda_{L,R},\,\Psi_{L,R}$ are given
by
\begin{eqnarray}
\Box A_\mu^{q} - k^2 z D_z^c \frac{1}{z} D_z^c A_\mu^{q} &=& 0,
\label{gauge-eom}
\\
\Box A_z^{q} - k^2 z D_z^c z D_z^c \frac{1}{z} A_z^{q} &=& 0,
\label{higgs-eom}
\\
\left\{
\not{\!\partial\,}
 - k \left( \frac{c_{\psi}}{z} + \gamma_5 D^c_z \right)
\right\}
e^{-2\sigma} \psi &=& 0,
\label{fermion-eom}
\end{eqnarray}
where 
$\not{\!\partial\,}=\eta^{\bar{m}\bar{n}}\gamma_{\bar{m}}\partial_{\bar{n}}$.
%
%
Gauge fields $A_M^{(a)}=(A_\mu^{(a)},A_z^{(a)}/z)$
and fundamental (adjoint) fermions $\psi^I_{L,R} = \Psi_{L,R}^{(i)}$ 
($\Lambda_{L,R}^{(a)}$) are decomposed into their Kaluza-Klein modes:
\begin{eqnarray}
A_M^{(a)}(x,z) &=& \sum_{n} f_{A_M,n}^{(a)}(z) A_{M,n}^{(a)},
\\
\psi^I_{L|R}(x,z) &=& e^{2\sigma} \sum_{n} f_{L|R,n}^I(z) \psi_{L|R,n}^I (x).
\end{eqnarray}
Here the left-(right-)handed fermions $\psi_{L(R)}$ are defined by
$\psi_{L(R)} = [1-(+)\gamma_5/2] \psi$. 
When $\langle A_z \rangle$ vanishes, or when an adjoint field
$\Phi_{\ad}^{(a)} = A^{q(a)}_M, \Lambda^{(a)}$ commute with the VEV of
gauge field: $ [\langle A_z^c \rangle, \Phi^{(a)}_M ] = 0$, we can
simplify Eqs. (\ref{gauge-eom}), (\ref{higgs-eom}), (\ref{fermion-eom}),
and we obtain wave equations for KK-mode functions as
\begin{eqnarray}
k^2 z \frac{d}{dz} \frac{1}{z} \frac{d}{dz} f_{A\mu,n}^{(a)}(z)
 &=&
- (m_{A\mu,n}^{(a)})^2 f_{A\mu,n}^{(a)}(z)
\label{modeeq-gauge}
\\
k^2\frac{1}{z} \frac{d}{dz} z \frac{d}{dz} \frac{f_{Az,n}^{(a)}(z)}{z}
 &=&
- (m_{Az,n}^{(a)})^2 \frac{f_{Az,n}^{(a)}(z)}{z}
\label{modeeq-higgs}
\\
k \left(\frac{c}{z} + \frac{d}{dz} \right) f_{\psi_R,n}^I(z)
&=&
m_{\psi,n}^I f_{\psi_L,n}^I(z)
,
\label{modeeq-l}
\\
k \left(\frac{c}{z} - \frac{d}{dz} \right) f_{\psi_L,n}^I(z)
&=&
m_{\psi,n}^I f_{\psi_R,n}^I(z)
.
\label{modeeq-r}
\end{eqnarray}
For a field $\Phi$ we obtain non-zero mode ($m_{\Phi} \neq 0$)
solutions for (\ref{modeeq-gauge}), (\ref{modeeq-higgs}),
(\ref{modeeq-l}), (\ref{modeeq-r}) into the form of
\begin{eqnarray}
f_{\Phi}(z) 
&=&
z^{s_{\Phi}}
\left[
 A_{\Phi} J_{\alpha_{\Phi}}(\hatm_{\Phi} z)
+
 B_{\Phi} Y_{\alpha_{\Phi}}(\hatm_{\Phi} z)
\right],
\label{modefunction0}
\end{eqnarray}
with $s_{\Phi} = \{1,1,\frac{1}{2} \}$, 
$\alpha_{\Phi} = \{1,0, \frac{1}{2} \pm c_{\psi} \}$, and
$\hatm_{\Phi}
\equiv m_{\Phi}/k$, for $\Phi = \{A_\mu,A_z, \psi_{R/L} \}$.
%
\par
At the two boundaries $y=y_{i}$ ($i=0,1$), we define the $Z_2$ boundary
conditions, namely, 
the odd boundary condition:
$
\Phi(x^\mu, y_i - y) = - \Phi(x^\mu, y_i + y)
$
and the even boundary condition:
$
\Phi(x^\mu, y_i - y) = + \Phi(x^\mu, y_i + y)
$, when $\Phi$ is the gauge boson or the ghost.
For fermions $\psi=\Psi,\Lambda$, we should write $Z_2$ boundary conditions as
\begin{eqnarray}
\psi(x^\mu, y_i - y) &=& \pm \eta_\psi \gamma_5 \psi(x^\mu, y_i + y),
\end{eqnarray}
where $\eta_{\psi}^2 = 1$.
For the KK mode functions $f_{\Phi,n}(y) \equiv f_{\Phi,n}(e^{\sigma(y)})$,
 the odd boundary conditions
can be written as
\begin{eqnarray}
\left. f_{\Phi,n}(y) \right|_{y_i} = 0 \quad (i=0,1),
\end{eqnarray}
whereas even boundary conditions take  the form of \cite{GP,Contino03}:
\begin{eqnarray}
\left.
\left(
 \frac{d f_{\Phi,n}(y)}{dy} + r_{\Phi}\sigma'(y) f_{\Phi,n}(y)
\right)
\right|_{y=y_i} =0 \quad (i=0,1),
\label{even-bc-0}
\end{eqnarray}
where $r_{\Phi} = 1$, $2$ and $\pm c$ for $\Phi=A_\mu$, $A_y$ and
$\psi_{R,L}$, respectively.
It would be convenient \cite{Hosotani06} to rescale the KK mode functions
(\ref{modefunction0}) and to define
\begin{eqnarray}
\check f_{\Phi,n}(z)
&\equiv&
z^{\check{\alpha}_{\Phi}} \left[
A_{\Phi,n} J_{\check{\alpha}_\Phi}(\hatm_{\Phi,n} z) + 
B_{\Phi,n} Y_{\check{\alpha}_\Phi}(\hatm_{\Phi,n} z)
\right]
\label{mode-function-rev}
\end{eqnarray}
where $\check{\alpha}_{\Phi} = 1,0,\text{ and }1/2 \pm c$
for $\Phi = A_\mu,\,A_z/z\text{ and }\psi_{R,L}$,
respectively.\footnote{We choose $\eta_{\psi}=+1$ for $\psi$.}
With (\ref{mode-function-rev}), we can rewrite even boundary conditions
(\ref{even-bc-0}) into the form of
 $\partial_z \check f_\Phi(z) = 0$.
%
\par
In this model, the extra-dimensional components of the gauge field $A_y$ can
develop a non-zero vacuum expectation value (VEV) $\langle A_y \rangle$.
Furthermore, since the non-simply connected topology of the extra
dimension of the RS space-time, we cannot gauge away the Wilson-line
phase :
$
\exp (i g\int_{y_0}^{y_1} dy \langle A_y \rangle)
$.
This phase can change the boundary condition and therefore can change
mass spectrum of fields due to Aharonov-Bohm effect \cite{Aharonov-Bohm}.
To see this, in the following we will consider one of the simplest but
non-trivial case.
In the beginning, we turned off gauge VEV and impose
a $Z_2$ boundary condition on the theory.
It is
\begin{eqnarray}
\tilde A_\mu(x,-y + y_{i}) = + \tilde P_{i} \tilde A_\mu \tilde P_{i}^{\dag},
\quad
\tilde A_y (x,-y + y_{i})  = - \tilde P_{i} \tilde A_y \tilde
 P_{i}^{\dag},
\label{AB-bc}
\end{eqnarray}
here $ \tilde P_{0} = \tilde P_{1} = \sigma_3$.
This boundary conditions (\ref{AB-bc}) arrows for $\tilde{A}_\mu^{3}$
and $\tilde{A}_y^{1,2}$ to have zero modes. Therefore this boundary
condition break the $SU(2)$ gauge symmetry to $U(1)_3$ with massless gauge
boson $\tilde{A}^{(3)}_\mu$.
Due to the non-simply connected nature of $S^1/Z_2$ topology of the extra
dimension of the RS space-time, $\tilde{A}_y^{(1,2)}$ can develop a VEV.
Without loss of generality, we can set $\langle \tilde{A}_y \rangle$ in the
direction of $A_y^{(2)}$.
As a zero-mode solution of (\ref{modeeq-higgs}), we set gauge VEV as
\begin{eqnarray}
\langle \tilde A_z(z) \rangle &=& v z \sigma_2,
\label{gauge-vev}
\end{eqnarray}
where $v$ is a constant parameter. 
The $U(1)_3$ gauge symmetry can be broken, if $v\neq0$ and if the Wilson
line phase $ \langle W \rangle = \exp ( i\int_{z_0}^{z_1} g \langle
\tilde{A}_z \rangle dz )$ does not commute with the generator of the
$U(1)_3$ : $\sigma_3/2$.
In the following we refer the gauge which is represented by non-zero
VEV (\ref{gauge-vev}) and $Z_2$ boundary condition (\ref{AB-bc}), as the
``Aharonov-Bohm gauge'' or the AB-gauge.
%
%
\par
By using a gauge transformation
\begin{eqnarray}
\tilde A_M &\to& A_M = 
\tilde\Omega \tilde A_M \tilde\Omega^{\dag}
 - \frac{i}{g} \tilde\Omega\partial_M \tilde\Omega^{\dag},
\\
\tilde \Omega(z) 
&=& 
\exp\left[
-ig \int_{z_0}^{z} d\zeta \langle \tilde A_z(\zeta) \rangle
\right]
=
\exp\left[-i\frac{gv}{2}(z^2-z_0^2)\sigma_2 \right],
\end{eqnarray}
we move to an another gauge $\langle A_z \rangle = 0 $.
The boundary condition of the gauge fields $\tilde P_{i}$ are also changed to
$P_{i}$:
\begin{eqnarray}
P_{0} = \tilde P_0 = \sigma_3,
\quad
P_{1} = e^{i\theta\sigma_2} \sigma_3 e^{-i\theta\sigma_2},
\label{twisted-bc}
\end{eqnarray}
by the gauge transformation:
$P_{i} = \tilde{\Omega}(z_i) \tilde P_{i} \, \tilde{\Omega}^\dag(z_i)$.
In Eq.(\ref{twisted-bc}), $\theta$ are given by
\begin{eqnarray}
\theta = \frac{gv}{2}\left(z_1^2 - z_0^2 \right).
\label{twisted-VEV}
\end{eqnarray}
Thus we find that in this gauge the gauge field $A_M^{(1,3)}$ obey
the twisted boundary condition (\ref{twisted-bc}) at $z=z_1$, whereas the
Wilson-line phase becomes trivial $\langle W \rangle = 1$.
We refer the gauge with vanishing gauge VEV and with the boundary
condition (\ref{twisted-bc}) as the (twisted) boundary condition gauge
or the BC-gauge.
%
\par
In the analogous way, we can set boundary conditions and perform gauge
transformations of fermions.
We can write the boundary conditions of $\Psi$ and $\Lambda$ in the
AB-gauge as\footnote{ For simplicity, we assume that all boundary
conditions take same form as gauge field, nevertheless the boundary
conditions of all fermions are not necessarily the same as the gauge
fields.}
\begin{eqnarray}
\tilde \Psi(-y + y_i) = \eta_\Psi \tilde P_{i} \tilde \Psi(+y+y_i),
&&
\tilde \Lambda(-y + y_i) =
 \eta_\Lambda \tilde P_{i} \tilde \Lambda(+y + y_i) \tilde P_{i}^{\dag},
\end{eqnarray}
with $\eta_{\Psi,\Lambda}^2=1$.
For simplicity, hereafter we set $\eta_{\Psi} = \eta_{\Lambda} = +1$.
%
\par
\begin{figure}
\caption{Plot of $x_{n=1,2,3}(\frac{1}{2}+c,\theta)$ (see text), for
 various $\theta$ and for $|c|=1/2$ (left), $|c|=1/4$ (center) and
 $|c|=0$ (right) with fixed $kR=12$.  The $n$-th zero $x_n$ of
 $N(\theta,x/z_1,\frac{1}{2}+c)$ corresponds to the $n$-th KK-mass $m_n$
 by the relation $m_n = k x_n/z_1$.}  \label{bc-mass}
 \includegraphics[width=0.28\linewidth]{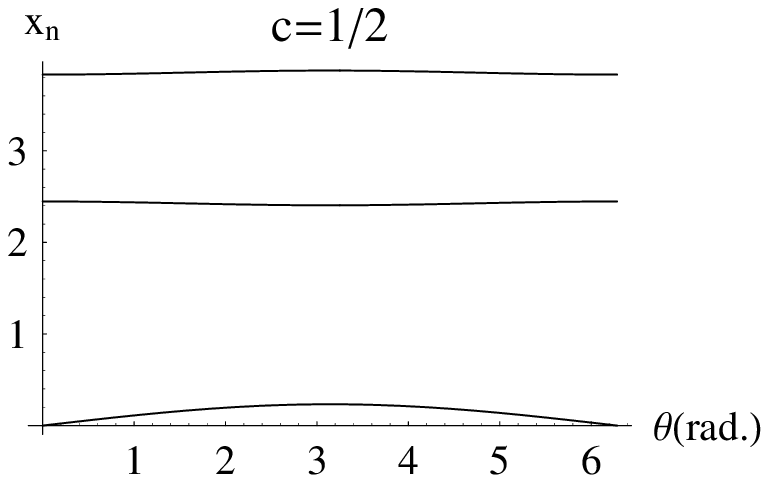}
 \includegraphics[width=0.28\linewidth]{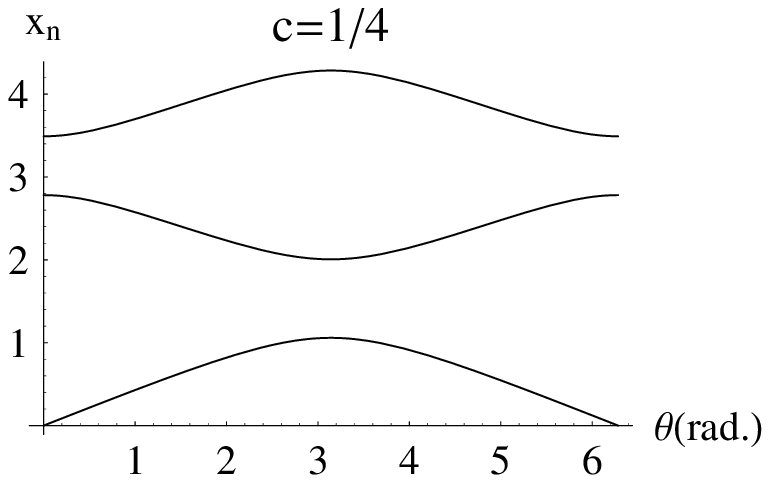}
 \includegraphics[width=0.28\linewidth]{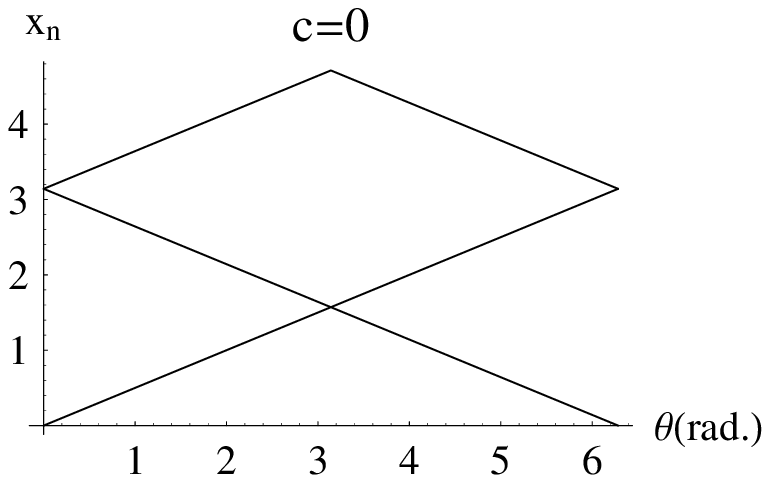}
\end{figure}

In the BC gauge, since the gauge VEV $\langle A_z \rangle$ vanishes, we
can solve the wave equations
(\ref{modeeq-gauge})(\ref{modeeq-higgs})(\ref{modeeq-l})(\ref{modeeq-r})
for $\Psi^{(i=1,2)}$, $A_M^{(i=1,3)}$ and $\Lambda^{(i=1,3)}$.
As an example, we consider the right-handed fundamental fermion
$\Psi^{(1,2)}_R$ with bulk mass $c$.
In the BC gauge, wave functions take the form of
\begin{eqnarray}
f^{(i)}_{R,n}(z) = z^{1/2} \left\{
a_{n}^{(i)} J_{c+1/2}(\hat m_n z) + b_{n}^{(i)} Y_{c+1/2}(\hat m_n z)
\right\}
\,\,\,
(i=1,2),
\end{eqnarray}
where $a_n$ and $b_n$ are constant, and we set
 $\hat{m}^{(1)}_n = \hat{m}^{(2)}_n \equiv \hat{m}_n$.
By solving twisted boundary condition in BC gauge
(or by going back to AB gauge by gauge transformation
$\Psi \to \tilde{\Psi} = \Omega \Psi$, with $\Omega =
(\tilde{\Omega})^{-1}$
and considering the even (odd) boundary
condition of $f^{(1)}_R$ ($f^{(2)}_R$) at $z=z_{0,1}$),
we obtain conditions:
\begin{eqnarray}
\begin{array}{lcl}
(-a^{(1)}  \sin \theta + a^{(2)} \cos \theta) J_{c+1/2}(\hat m_n z_i) + 
(-b^{(1)}  \sin \theta + b^{(2)} \cos \theta) Y_{c+1/2}(\hat m_n z_i) &=& 0,
\\
(a^{(1)} \cos \theta + a^{(2)} \sin \theta) J_{c-1/2}(\hat m_n z_i) +
(b^{(1)} \cos \theta + a^{(2)} \sin \theta) Y_{c-1/2}(\hat m_n z_i) &=& 0,
\\
 && (i=0,1).
\end{array}
\end{eqnarray}
In the same way we can write the boundary condition of $\Psi_L^{(1,2)}$,
$A_M^{(1,3)}$ and $\Lambda_{L,R}^{(1,3)}$.
These boundary condition can be written in the form of
\begin{eqnarray}
M_{\Phi}(\hatm) V_{\Phi} &=& 0,
\end{eqnarray}
where 
$V_{\Phi} = (a_{\Phi}^{(3)},b_{\Phi}^{(3)},a_{\Phi}^{(1)},b_{\Phi}^{(1)})$ 
for $\Phi = A_M^{(3,1)},\Lambda^{(3,1)}_{L,R}$,
and
$V_{\Phi} = (a_{\Phi}^{(1)},b_{\Phi}^{(1)},a_{\Phi}^{(2)},b_{\Phi}^{(2)})$ 
for $\Phi = \Psi^{(1,2)}_{L,R}$.
The matrix $M_{\Phi}$ is defined by
\begin{eqnarray}
M_{\Phi}(\hatm)
&=&
\begin{pmatrix}
-s_1 J_{\alpha}(\hatm z_1) &
-s_1 Y_{\alpha}(\hatm z_1) &
 c_1 J_{\alpha}(\hatm z_1) &
 c_1 Y_{\alpha}(\hatm z_1) 
\\
-s_0 J_{\alpha}(\hatm z_0) &
-s_0 Y_{\alpha}(\hatm z_0) &
 c_0 J_{\alpha}(\hatm z_0) &
 c_0 Y_{\alpha}(\hatm z_0) 
\\
 c_1 J_{\alpha-1}(\hatm z_1) &
 c_1 Y_{\alpha-1}(\hatm z_1) &
 s_1 J_{\alpha-1}(\hatm z_1) &
 s_1 Y_{\alpha-1}(\hatm z_1) 
\\
 c_0 J_{\alpha-1}(\hatm z_0) &
 c_0 Y_{\alpha-1}(\hatm z_0) &
 s_0 J_{\alpha-1}(\hatm z_0) &
 s_0 Y_{\alpha-1}(\hatm z_0) 
\end{pmatrix},
\end{eqnarray}
where
$\alpha = 1, 0, c+1/2\text{ and }c-1/2$ for $A_\mu$, $A_z$,
$\psi_R\text{ and }\psi_L$ ($\eta_\psi = +1$), respectively.
$s_i$ and $c_i$ are defined by
\begin{eqnarray}
s_i(c_i)
 = \sin(\cos) \left(\frac{n_\Phi \theta}{2}\frac{z_i^2-z_0^2}{z_1^2-z_0^2}
 + \delta_\Phi \right),
\end{eqnarray}
where
$\delta_\Phi = 0$      for $\Phi=A_\mu,\Psi_R,\Lambda_{R}$, and
$\delta_\Phi = -\pi/2$ for $\Phi=A_z,\Psi_L,\Lambda_{L}$.
The factor $n_{\Phi}$ depends on $\Phi$'s
representation of the gauge group.  In $SU(2)$ case, $n_\Phi=1\,(2)$ when
$\Phi$ is in fundamental (adjoint) representation.
The $n$-th Kaluza-Klein state of the field $\Phi$ has a mass $m_{\Phi,n}
= k \hatm_{\Phi,n}$, where the $\hatm_{\Phi,n}$ are defined as a $n$-th
smallest solution of the equation:
$
\det M_\Phi(\hatm) 
= 0
$.
Defining
\begin{eqnarray}
N(\theta,\hatm,\alpha)
&\equiv&
\frac{2\cos (n_{\Phi}\theta)}{\pi^2}
\nonumber\\&&
+ \frac{1}{2} \hatm^2 z_1 (
F_{\alpha,\alpha-1}(\hatm, z_1) F_{\alpha-1,\alpha}(\hatm, z_1) +
F_{\alpha,\alpha}(\hatm, z_1) F_{\alpha-1,\alpha-1}(\hatm, z_1)
)
\end{eqnarray}
(where 
$
F_{\alpha,\beta}(\hatm,z) 
\equiv J_{\alpha}(\hatm z) Y_{\beta}(\hatm) -
J_{\beta}(\hatm) Y_{\alpha}(\hatm z)
$
), we rewrite the condition $ \det M_\Phi(\hatm) = 0 $ as
\begin{eqnarray}
N(n_{\Phi}\theta,\hatm,\alpha_\Phi) = 0,
\label{det=0}
\end{eqnarray}
where $\alpha_\Phi$ and $n_\Phi$ are given by
\begin{eqnarray}
\begin{tabular}{lcc}
\hline\hline
$\Phi$ & $\alpha_\Phi$ & $n_{\Phi}$ 
\\
\hline
$A_M^{(1\leftrightarrow3)}$ & $1$ & $2$
\\
$A_M^{(2)}$ & $1$ & $0$
\\
$\Psi^{(1\leftrightarrow2)}$ 
& $\frac{1}{2}+c_{\Psi}$ & 1
\\
$\Lambda^{(1\leftrightarrow3)}$ 
& $\frac{1}{2} + c_{\Lambda}$ & 2
\\
$\Lambda^{(2)}$ 
& $\frac{1}{2} + c_{\Lambda}$ & 0
\\
\hline\hline
\end{tabular}.
\end{eqnarray}
Let $x_n(\alpha,\theta)$ be the $n$-th smallest non-negative zeros of
$N(\theta,x/z_1,\alpha)$.
Thus, the $n$-th Kaluza-Klein mass of the field $\Phi$ : $m_{\Phi,n}$ is
given by
\begin{eqnarray}
m_{\Phi,n} = (k/z_1)\cdot x_n(\alpha_{\Phi},n_{\Phi}\theta).
\end{eqnarray}
In FIG. \ref{bc-mass} we plot $x_i(\frac{1}{2}+c,\theta)$ ($i=1,2,3$)
which are first three smallest zeros of $ N(\theta, x/z_1,
\frac{1}{2}+c)$ with $kR = 12$.
We see that the mass spectrum largely depends
on $\theta$ when $|c|$ is small, whereas the dependence on $\theta$
become smaller for larger $|c|$ (here we note that $N(\theta,x,\alpha)$
has a symmetry 
$ N(\theta,x,\alpha) = N(\theta,x,1-\alpha)$,
as shown in Appendix B of \cite{Hosotani06}).%
\footnote{ FIG.~\ref{bc-mass} of the present paper and FIG.~1 in
\cite{Hosotani06} are similar to each other, although in each
figure different parameters are changed ($|c|$ in the former and $k R$ in
the latter.)}
From the dependence of $x_n(\alpha,\theta)$ as we have seen above, we
can expect that the the effective potential, which is obtained by
summing up all of the Kaluza-Klein mode, has large (small) dependence on
$\theta$ when $|c|$ is small (large).
%
%
%
\par
\begin{figure}
\caption{Plot of $v(\theta,1/2+c)$ with $|c|=0.0$(thin solid),
 $|c|=0.4$(dashed) and $|c|=0.5$(thick solid).}\label{thetaplot}
\includegraphics[width=0.6\linewidth]{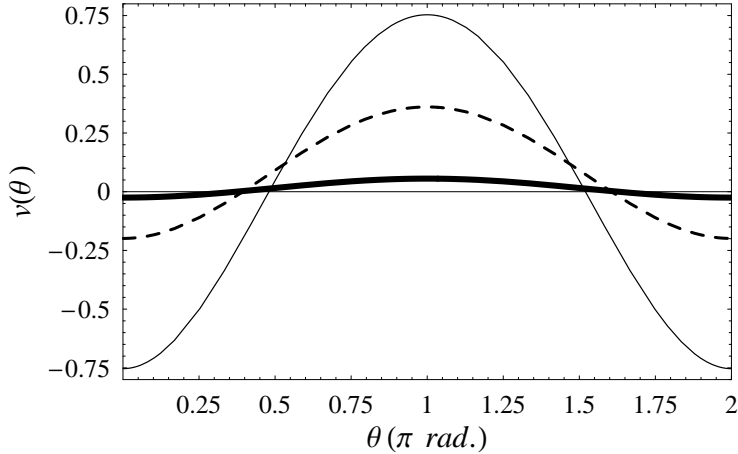}
\end{figure}
\begin{figure}
\caption{Plot of $\Delta v(1/2+c)\equiv
v(\pi,1/2+c)-v(0,1/2+c)$ with $a = \exp(-12\pi)$.}\label{plotv}
\includegraphics[width=0.6\linewidth]{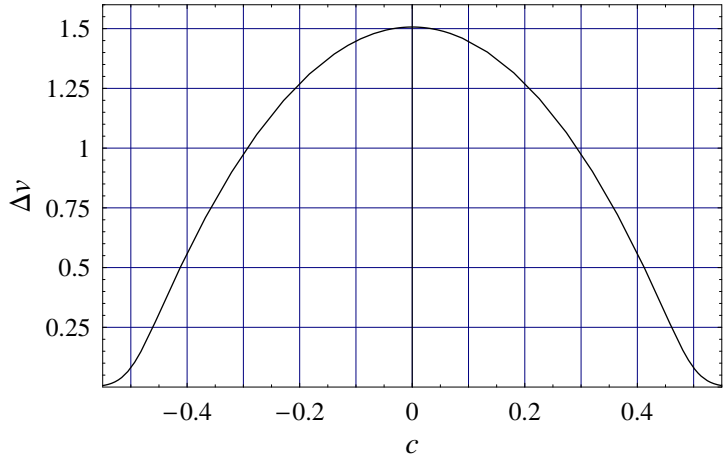}
\end{figure}
\begin{figure}
\caption{%
The ratio $v_r(\theta,c)/(x_1(\frac{1}{2}-c,\theta))^2$ (see text) with
 $kR=12$, for $c=0.5$ (thick solid), $0.4$ (thin solid), $0.2$ (thin dashed)
 and $0.0$ (thick dashed), respectively.}\label{fig-vmratio}
 \includegraphics{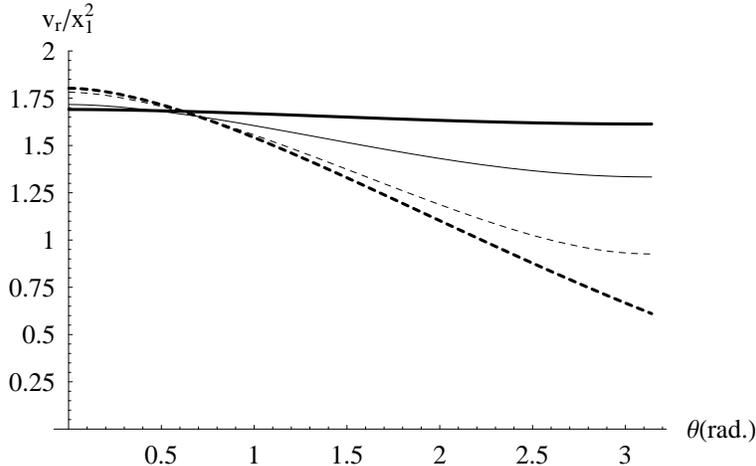}
\end{figure}

We evaluate the 1-loop effective potential in a similar way as \cite{CW}.
As an example we calculate the one-loop contribution of a fundamental fermion
$\Psi$. This is given by
\begin{eqnarray}
-4 I = -4\cdot \frac{-\mu^{4-d}}{2}
\int \frac{d^d p}{i(2\pi)^d}
\sum_n \log(-p^2 +m_n^2),
\end{eqnarray}
where $d$ defined as $d=4-\epsilon$, $\mu$ is the renormalization scale,
and $m_n$ is the $n$-th positive-real solution of
$N(\theta,m/k,\frac{1}{2}+c_\Psi)=0$.
The coefficient $-4$ reflects the four degrees of freedom of a Dirac
fermion and the minus sign for the fermion loop.
After performing the dimensional regularization, we can write 
$I$ as
\begin{eqnarray}
I = \frac{\mu^{\epsilon}}{2}\cdot
\frac{\Gamma(-2+\epsilon/2)}{(4\pi)^{2-\epsilon/2}}
\sum_{n} m_n^{4-\epsilon}.
\end{eqnarray}
Here the summation $\sum_n m_n^{-s}$ can be rewritten as a contour
integral:
\begin{eqnarray}
\sum_n m_n^{-s} = 
 \frac{k^{-s}}{2\pi i} \oint_C dz\, 
 z^{-s}
 \frac{N'(\theta,z,\frac{1}{2}+c_{\Psi})}{N(\theta,z,\frac{1}{2}+c_{\Psi})},
\label{contour}
\end{eqnarray}
where the path $C$ is a set of all circles surrounding each zeros of
$N(\theta,z,\frac{1}{2}+c_\Psi)$ on the positive real axis. 
The contour integral in (\ref{contour}) can be calculated in a similar
way to \cite{Toms,Oda,RSquantum}.
We find that $\theta$-dependent part of $I$ is finite.
\begin{eqnarray}
I = \frac{1}{32\pi^2} 
\left( \frac{k}{z_1} \right)^4 v(\theta,\tfrac{1}{2} + c_\Psi)
 + \text{($\theta$-independent)}
,
\end{eqnarray}
where $v(\theta,\nu)$ is defined by
\begin{eqnarray}
v(\theta,\nu) 
&\equiv&
\int_0^{\infty} dx \, x^3 \log \left[
1 + \frac{1}{2}
\left(
\frac{I_{\nu  }(ax) K_{\nu-1}( x)}{
      I_{\nu-1}( x) K_{\nu  }(ax)}
-
\frac{I_{\nu  }(ax) K_{\nu  }( x)}{
      I_{\nu  }( x) K_{\nu  }(ax)}
\right.\right.
\nonumber\\&&
\left.\left.
-
\frac{I_{\nu-1}(ax) K_{\nu-1}( x)}{
      I_{\nu-1}( x) K_{\nu-1}(ax)}
+
\frac{I_{\nu-1}(ax) K_{\nu  }( x)}{
      I_{\nu  }( x) K_{\nu-1}(ax)}
\right)
\right.
\nonumber\\&&
\left.
+ 
\frac{I_{\nu-1}(ax) I_{\nu}(ax) K_{\nu-1}( x) K_{\nu}( x)}{
      I_{\nu-1}( x) I_{\nu}( x) K_{\nu-1}(ax) K_{\nu}(ax)}
- \frac{\cos\theta}{2 a x^2 
      I_{\nu-1}( x) I_{\nu}( x) K_{\nu-1}(ax) K_{\nu}(ax)}
\right],
\label{anIntegral}
\end{eqnarray}
with $a \equiv 1/z_1$.
This function $v(\theta,\nu)$ has a symmetry 
$v(\theta,\nu) = v(\theta,1-\nu)$,
and a periodicity $V(\theta,\nu) = v(\theta+2\pi,\nu)$.
The function (\ref{anIntegral}) with $\nu=1$ becomes the
same form as the integral in Eq.~(18) of \cite{Oda}.
Furthermore, when $\nu=1/2$ we can rewrite $v(\theta,1/2)$  as
\begin{eqnarray}
v(\theta,1/2) 
&=& 
\left(\frac{1}{1-a}\right)^4
\int_0^{\infty} \! dt\, t^3 \log
\left[1 - \frac{\cos\theta}{\cosh 2t}\right]
+\text{($\theta$-independent)}
\nonumber\\
&=&
\left(1-a\right)^{-4}
\left\{
-\tfrac{3}{4} \Re \left[ \Li_5(e^{i\theta}) \right]
- \tfrac{45}{2048}\zeta(5)
\right\}
+ \text{($\theta$-independent)}
,
\label{bulkless}
\end{eqnarray}
where $\Li_d(x)$ is the polylogarithm and
$\zeta(x)$ is the Riemann zeta function.
$\Re\,\Li_d(e^{i\theta})$ can be expanded as $\Re\,\Li_d(e^{i\theta}) =
\sum_{n=1}^{\infty} \cos n\theta/n^d$, or expanded as
$ \Re\,\Li_5(e^{i x}) = \zeta(5) - \frac{1}{2}\zeta(3) x^2 +
\left(\frac{25}{288}- \frac{1}{24}\log x \right) x^4 + {\cal O}(x^6)$.
The expression (\ref{bulkless}) coincides with the result obtained in
Sec. IV of Ref. \cite{Toms}.
The shapes of $v(\theta,\nu)$ with respect to $\theta$ with different $\nu$
are similar with each other (FIG. \ref{thetaplot}).
%
We show dependence of amplitude $\Delta v(\frac{1}{2}+c) \equiv v(\pi,\frac{1}{2}+c) -
v(0,\frac{1}{2}+c)$ on $c$ in FIG.~\ref{plotv}.
%
In the case of flat extra dimension it is known \cite{HIL,bulkmass} that
the contribution to one-loop effective potential from a field dumps when
the bulk mass of the field is much larger compared to (the inverse of)
the size of the extra dimension.
It is also true for the models in the RS space-time.
%
%
However, in the RS space-time the relation between the effective
potential and the 4-dimensional mass of the field is different from the
case of flat extra dimension. Therefore a fermion with the large lowest
KK mass can have a large contribution to the effective potential.
To see this, let us see the relation between the magnitude of the
effective potential and the mass of the fermion at the lowest KK level.
The lowest KK mass $m(c_\Psi)$ of a $SU(2)$ fundamental fermion with
bulk mass $c_{\Psi}$ is given by
$
m(c_\Psi) = (k/z_1)\cdot x_1(\frac{1}{2}\pm c_\Psi, \theta).
$
And we define the magnitude of the one-loop effective potential from the
 fermion as $v_r(\theta,c) \equiv v(\theta,c) - v(0,c)$.
In FIG.~\ref{fig-vmratio}, we show the relation between $v_r(\theta,c)$
and the square of $x_1(\frac{1}{2}-c,\theta)$.
From FIG.~\ref{fig-vmratio}, it can be safely said that the magnitude of
the one-loop contribution to the effective potential is proportional to
the square of the lowest KK mass of the fermion when the absolute value
of bulk mass of the fermion is sufficiently large.
This means that when the fermion's lowest-KK mass is small (large) the
contribution of the fermion to the effective potential is small (large).
%
\par
Now we are ready to discuss the gauge symmetry breaking by Wilson line
phase.
The effective potential of the $SU(2)$ model consists of gauge and ghost
loop contribution $V_{\gh}$, contribution from the fundamental
fermion $\Psi$ $V_{\fd}$ and/or in adjoint fermion $\Lambda$: $V_{\ad}$.
They are given by ($\theta$-independent part is omitted.),
\begin{eqnarray}
\begin{array}{lcl}
V_{\gh}(\theta) &=& \phantom{-} 3\C \cdot v(2\theta,1),
\\
V_{\fd}(\theta,c_{\Psi}) &=& -4\C \cdot v(\theta,\frac{1}{2} + c_{\Psi}),
\\
V_{\ad}(\theta,c_{\Lambda}) &=& -4\C \cdot v(2\theta, \frac{1}{2} + c_{\Lambda}),
\end{array}  
\end{eqnarray}
where $\C \equiv (k/z_1)^4/ 32\pi^2$.
%
%
In the AB gauge, the $SU(2)$ gauge symmetry is at first broken by the
boundary conditions $\tilde{P}_{0,1}$.
The generator $T^{(3)}$ commutes with both of the boundary conditions and
becomes the generator of the $U(1)_3$ gauge symmetry.
When the Wilson-line phase becomes non-trivial and does not commutes
with $T^{(3)}$, the remaining $U(1)_3$ gauge symmetry could be broken.
The Wilson-line phase is determined dynamically as a configuration
which minimize the effective potential.
As a first example, we consider the model in which the gauge field and a
fundamental fermion is included.
The effective potential in this case can be written as $ V_{\gh}(\theta)
+ V_{\fd}(\theta,c_{\Psi}) $.
This potential has a global minimum at $\theta = \pi$ ($0 \le
\theta \le \pi$) for any value of $c_{\Psi}$.
When $\theta=\pi$, the $U(1)_3$ gauge symmetry remains unbroken because the
Wilson-line phase
$
 \langle W(\theta=\pi) \rangle
  = \exp \left[i\pi\sigma_2\right] = \diag(-1,-1)
$
can commute with the $U(1)_3$ generator $T^{(3)}$.
As a next example, we consider the model in which the gauge field and an
adjoint fermion with bulk mass $c_\Lambda$ are included.
We can write the effective potential as
$V_\gh(\theta) + V_{\ad}(\theta,c_{\Lambda})$. 
This has global minima at $\theta = 0,\pi$
($0 \le \theta \le \pi $) for $|c_{\Lambda}| > 0.507$,
and has a global minimum at $ \theta = \theta^{\min} \sim \pi/2$ for
$|c_{\Lambda}| < 0.507$.
In the latter case the the non-trivial Wilson line
$
\langle W(\theta^{\min}) \rangle 
\sim \exp\left[i(\pi/2) \sigma_2 \right] = i \sigma_1
$
cannot commute with $T^{(3)}$.
Thus the $U(1)_3$ gauge symmetry can be broken by the Wilson line, as seen
in the case of flat-extra dimension \cite{adjoint-fermion,Hosotani89}.
And the pattern of the symmetry-breaking depends on the bulk mass of the
fermion, as seen in \cite{bulkmass} in the case of the flat extra
dimension.
%
\section{SU(3) models}\label{sec-su3model}
%
In this section we consider models with the $SU(3)_w$ gauge symmetry in
the RS as an extension of electroweak theory.
\subsection{$SU(3)_w$ model}
In this model, the electroweak $SU(2) \times U(1)$ gauge symmetry is
enlarged to the $SU(3)_w$ gauge symmetry.
We define generators of $SU(3)_w$ as $T^a \equiv \lambda_{a}/2$
($a=1,...,8$), where $\lambda_{a}$ are the Gell-Mann matrices.
We also define $T^{9} \equiv \lambda_9/2$ and $T^{10}\equiv
\lambda_{10}/2$ ($\lambda_9 \equiv \diag(0, 1, -1)$ and $\lambda_{10}
\equiv \diag(-2,1,1)/\sqrt{3}$) as generators of $U(1)_{9}$ and
$U(1)_{10}$ gauge symmetries, respectively.
Following the way in
\cite{Antoniadis-Benakli-Quiros,Lim-Kubo-Yamashita,Hosotani04,Hosotani06},
we introduce $SU(3)_w$-triplet fermions, namely ``quarks''
$\Psi_{\fd={\rm u,c,t}}$, and ``leptons'' $\Psi_{\fd={\rm e},\mu,\tau}$:
\begin{eqnarray}
\Psi_{\rm u(c,t)}^{(c=1,2,3)} = 
({\rm d}({\rm s},{\rm b})_l,\, {\rm u}({\rm c},{\rm t})_l,\, {\rm
u}({\rm c},{\rm t})_r)^T,
&&
\Psi_{{\rm e}(\mu,\tau)} = 
(\nu_{{\rm e}(\mu,\tau),l},\, {\rm e}(\mu,\tau)_l,\, {\rm e}(\mu,\tau)_r)^T,
\end{eqnarray}
where the superscript $(c)$ denotes the $SU(3)_{\rm
color}$-charge.
For simplicity, we introduce the bulk mass terms of $\Psi_{\fd}$ fields
in the diagonal form, i.e., the mass terms for fermions in the action is
given by
\begin{eqnarray}
S_{\Psi,{\rm mass}}
= - \sum_{\fd} \int \!\! d^4x \!\! \int \!\! dy |E|
  c_{\fd} \epsilon(y)k \bar{\Psi}_{\fd} \Psi_{\fd},
\end{eqnarray}
in $SU(3)_w$ gauge basis.
$c_{\fd}$ is the bulk mass parameter of the fermion $\Psi_{\fd}$.
If the mass matrix is not diagonal, various phenomena of the
flavor-mixing will be observed \cite{flavor_on_RS}.
\par
In order to break the $SU(3)_w$ to the ``electroweak'' $SU(2) \times
U(1)$ symmetry, and to obtain chiral fermion zero modes, we chose the
orbifold boundary condition in the AB-gauge as
\begin{eqnarray}
\tilde{P}_i = \diag(-1,-1,+1) \quad (i = 0,1).
\label{su3bc}
\end{eqnarray}
When we turn off the gauge VEV $\langle A_z \rangle = 0$, the
$SU(3)_w$ symmetry is broken to $SU(2)_w \times U(1)_8$ only by the
boundary condition (\ref{su3bc}), where the unbroken generators of
$SU(2)_w$ and $U(1)_8$ are $T^{(1,2,3)}$ and $T^{(8)}$, respectively.
Gauge fields $A_\mu^{(1,2,3,8)}$ and $A_z^{(4,5,6,7)}$ have massless
modes.
Zero modes of the fermions $\Psi_{{\rm u},0}$ and $\Psi_{{\rm e},0}$
are given by\footnote{We choose $\eta_\Psi = +1$.}
\begin{eqnarray}
\Psi_{{\rm u},0} = 
({\rm d}^{\rm C}_{l,L},\, {\rm u}^{\rm C}_{l,L},\, {\rm u}^{\rm C}_{r,R} )^T,
\quad
\Psi_{{\rm e},0} = 
({\rm \nu}_{l,L},\, {\rm e}_{l,L},\, {\rm e}_{r,R} )^T,
\end{eqnarray}
where the superscript ${\rm C}$ denotes the charge conjugation (here we
omitted indices for the color charge).
%
%
The zero modes of $A_z^{(4,5,6,7)}$ can develop a VEV.
Here we assume that $A_z$ develops a VEV in the direction of
$\lambda_7$.
In the AB-gauge, we set
\begin{eqnarray}
\langle \tilde{A}_z \rangle 
&=& \frac{2\thetaH}{g}\frac{z}{z_1^2-z_0^2} 
\begin{pmatrix}
\phantom{0}&\phantom{0}&\phantom{0}\\ &&-i\\&\phantom{-}i&\end{pmatrix},
\end{eqnarray}
where $\thetaH$ is the Wilson-line phase parameter which will be determined
dynamically.
Once the gauge VEV is turned on : $\thetaH \neq 0$, the $SU(2)_w$ symmetry
is broken, and ``up type quarks'' ${\rm u},{\rm c},{\rm t}$ and
``charged leptons'' ${\rm e},\mu,\tau$ get each mass terms through the
gauge coupling, i.e., $\bar\Psi_u g \tilde{A}_z^{(7)} \Psi_u$.
%
%
We can move to the BC-gauge by the singular gauge transformation with
$\Omega = \exp(i\thetaH\lambda_7)$.
This gauge transformation reads
\begin{eqnarray}
\begin{pmatrix} 
\tilde{A}_M^{(1)}-i\tilde{A}_M^{(2)} \\ \tilde{A}_M^{(4)} - i\tilde{A}_M^{(5)}
\end{pmatrix}
&=&
\begin{pmatrix}
  \cos\thetaH & \sin\thetaH \\
 -\sin\thetaH & \cos\thetaH
\end{pmatrix} 
\begin{pmatrix}
A_M^{(1)}-iA_M^{(2)} \\ A_M^{(4)} - iA_M^{(5)}
\end{pmatrix},
\\
\begin{pmatrix} 
\tilde{A}_M^{(9)} \\ \tilde{A}_M^{(6)}
\end{pmatrix}
&=&
\begin{pmatrix}
  \cos2\thetaH & \sin2\thetaH \\
 -\sin2\thetaH & \cos2\thetaH
\end{pmatrix} 
\begin{pmatrix}
A_M^{(9)} \\ A_M^{(6)}
\end{pmatrix},
\\
\begin{pmatrix} 
\tilde{\Psi}^{(2)} \\ \tilde{\Psi}^{(3)}
\end{pmatrix}
&=&
\begin{pmatrix}
  \cos \thetaH & \sin \thetaH \\
 -\sin \thetaH & \cos \thetaH
\end{pmatrix} 
\begin{pmatrix}
\Psi^{(2)} \\ \Psi^{(3)}
\end{pmatrix},
\end{eqnarray}
Thus under the gauge transformation $\Omega$, $A_M^{(1,4)}$, $A_M^{(2,5)}$
and $\Psi^{(2,3)}$ transform like a $SU(2)$ fundamental field
respectively, whereas $A_M^{(9,6)}$ transforms like $SU(2)$ adjoint
field in the previous section.
%
Each pair of the gauge field zero modes
$(A_{\mu,0}^{(i)},A_{z,0}^{(j)})$ ($(i,j) = (1,4),(2,5),(9,6))$) and
fermion zero modes $(\Psi_{\fd,L,0}^{(2)},\Psi_{\fd,R,0}^{(3)})$ can
obtain mass terms through the gauge coupling.
We define these massive field as $A_M^{(1\leftrightarrow4)}$,
$A_M^{(2\leftrightarrow5)}$, $A_M^{(6\leftrightarrow9)}$ and
$\Psi^{(2\leftrightarrow3)}$, respectively.
Zero modes of $A_\mu^{(10)}$, $A_z^{(7)}$ and $\Psi^{(1)}_{\fd,0,L}$
remain massless. Especially, $A_{\mu,0}^{(10)}$ and $A_z^{(7)}$ are
identified with the $U(1)_{10}$ gauge field (``photon'') and one of the
$SU(2)_w$-doublet ``Higgs'', respectively. These fields are massless at
tree level.
%
\par
The $n$-th KK masses $m_{\Phi,n}$ of a field $\Phi$ are
given by
$
m_{\Phi,n} = (k/z_1)\cdot x_{n}(\alpha_\Phi, n_{\Phi} \thetaH)
$,
where $\alpha_\Phi$ and $n_\Phi$ for a field $\Phi$ are summarized in
(\ref{su3spectrum-1}).
\begin{eqnarray}
\begin{tabular}{lcc}
\hline\hline
$\Phi$ & $\alpha_\Phi$ & $n_{\Phi}$ \\
\hline
$A_M^{(1\leftrightarrow4,2\leftrightarrow5)}$ & $1$ & $1$ 
\\
$A_M^{(6\leftrightarrow9)}$ & $1$ & $2$ 
\\
$A_M^{(7,10)}$ & $1$ & $0$ 
\\
$\Psi_\fd^{(2\leftrightarrow3)}$ & $\frac{1}{2} + c_{\fd}$ & $1$ 
\\
$\Psi_\fd^{(1)}$ & $\frac{1}{2} + c_{\fd}$ & $0$ 
\\
\hline\hline
\end{tabular}
\label{su3spectrum-1}
\end{eqnarray}
Now we discuss about the fermion masses in the 4-dimensional effective
theory. We define the function of the lowest KK mass of a field $\Phi$
$\mu(\alpha_\Phi,\thetaH,n_\Phi)$ as
\begin{eqnarray}
\mu(\alpha_\Phi,\thetaH,n_\Phi) 
\equiv 
\frac{k}{z_1} x_{1}(\alpha_\Phi, n_{\Phi} \thetaH).
\end{eqnarray}
As pointed out in \cite{Hosotani06}, it is useful to write the mass of
the fields in unit of the W-boson mass $m_W$. We rewrite
$\mu(\alpha,theta,n))$ as
\begin{eqnarray}
\mu_W(\alpha_\Phi,\thetaH,n_\Phi)
\equiv m_W \cdot \frac{\mu(\alpha_\Phi,\thetaH,n_\Phi)}{\mu(1,\thetaH,1)}
= m_W \cdot \frac{x_1(\alpha_\Phi,n_\Phi
\thetaH)}{x_1(1,\thetaH)},
\end{eqnarray}
here we have used $m_W = \mu(1,\thetaH,1)$.
With these quantity we can write the lowest KK-mass of fields.
For example, masses of ``quark'' and ``lepton''
$m_{\fd={\rm u},{\rm c},{\rm t},{\rm e},\mu,\tau}$ can be written as 
$ m_{\fd} = \mu_W(\frac{1}{2}+ c_\fd,\thetaH,1) $.  
\par
If the mass of a field is sufficiently smaller than $m_{KK} \equiv \pi
k/(z_1-1)$, we can utilize approximation formulas in
Appendix \ref{sec-approx}.
For $kR = 12$, $\pi m_{W}/m_{KK} \approx x_1(1,\theta) \gtrsim 0.2326$.
In the region $0 \le \thetaH \le \pi$, $m_{KK}$ is a monotonically
decreasing function of $\thetaH$, and we find the lower bound :
\begin{eqnarray}
m_{KK} \ge m_{KK}(\thetaH=\pi) \simeq 1086\text{\GeV}.
\label{mkk-bound}
\end{eqnarray}
%
%
Since quark and lepton masses are all sufficiently smaller than
$m_{KK}$'s lower bound (\ref{mkk-bound}), we can use approximation
formulas in Appendix \ref{sec-approx} to rewrite the mass of a
fundamental fermion with bulk mass $c_{\fd}$.
It is written in the form of
\begin{eqnarray}
m_{\fd} 
\simeq 
\mubar_W(\tfrac{1}{2} + c_{\fd},\thetaH,1)
=
m_W
\sqrt{
\frac{z_1 (c_{\fd}+\frac{1}{2})(c_{\fd}-\frac{1}{2})kR\pi}{
2 \sinh[(c_{\fd}+\frac{1}{2}) kR\pi]\sinh[(c_{\fd}-\frac{1}{2})kR\pi]}
}.
\label{fund-mass}
\end{eqnarray}
We see from (\ref{fund-mass}) that masses of ``quarks'' and ``leptons''
are almost independent of $\thetaH$.
Therefore we can find values of bulk mass $c_{\fd}$ for each ``quarks''
and ``leptons'' from its 4D mass irrespective of the value of
$\thetaH$ which is obtained by solving the Wilson-line dynamics.
Thus, from the formula (\ref{fund-mass}), we obtain bulk masses of ``top
quark'' $c_{{\rm top}}$ and other ``quarks'' and
``leptons'' \cite{Hosotani06}.  We obtain $c_{{\rm top}}\simeq 0.4366$
and $|c_{{\rm u},{\rm c},{\rm e},\mu,\tau}| > 0.6$, respectively.
%
%
%
In the way similar to
\cite{Antoniadis-Benakli-Quiros,Lim-Kubo-Yamashita,Hosotani04} but
extended for the warped space-time, we obtain the effective potential:
\begin{eqnarray}
V_{\eff}^{\gh+\mathrm{t}} &=& V_{{\rm gauge}} + V_{\mathrm{top}},
\\
V_{{\rm gauge}} &=& + 3\C \cdot \left[ 2v(\thetaH,1) + v(2\thetaH,1)\right],
\\
V_{\mathrm{top}} &=& -3\cdot4\C \cdot v(\thetaH,\tfrac{1}{2}+c_{\mathrm{top}}),
\end{eqnarray}
where we have neglected contributions from other quarks and leptons,
because they have small contributions to the effective potential with
their large bulk masses.
The factor $3$ in $V_{\mathrm{top}}$ reflects the degrees of freedom
of $SU(3)_{\rm color}$-color charge.
The effective potential $V_{\eff}^{\gh+\mathrm{t}}$ has the global
minimum at $\thetaH = \pi $.
The Wilson-line phase with $\thetaH=\pi$ (in AB-gauge) is
\begin{eqnarray}
\langle W(\thetaH=\pi) \rangle 
= \exp \left[i \pi \lambda_7 \right] = \diag(1, -1, -1),
\end{eqnarray}
and it can commute with both of boundary condition $\tilde{P}_{0,1}$ and
with both of two $U(1)$ generators $T^{(10)}$ and $T^{(9)}$
simultaneously.
Hence the $SU(2)_w \times U(1)_8$ symmetry is broken to $U(1)_{9}\times
U(1)_{10}$. Besides the ``electro-magnetic'' $U(1)_{10}$, one extra
$U(1)_{9}$ symmetry remains unbroken.
%
\par
When the effective potential $V_{\eff}(\thetaH)$ has the global minimum at
$\thetaH =\thetaH^{\min}$, the one-loop Higgs mass $m_h$ is written in
terms of $m_W$ as
\begin{eqnarray}
m_h &\simeq& m_W \frac{kR}{4} 
\sqrt{\pi \alpha_W V_{\eff}^{(2)}(\thetaH^{\min})/\C} 
\left| \csc\frac{\thetaH^{\min}}{2} \right|,
\label{higgs-mass}
\end{eqnarray}
where $\alpha_W$ is 4D fine-structure constant of $SU(2)$ before the
symmetry breaking\footnote{The 4-dimensional coupling is given by $g_4^2
= g^2/\pi R$.},
and $V^{(2)}_{\eff}(\thetaH^{\min}) \equiv \partial^2
V_{\eff}(\thetaH)/\partial\thetaH^2|_{\thetaH=\thetaH^{\min}}$.
By substituting $m_W=80.4\,\GeV$, $\alpha_W=0.032$, $V_{\eff} =
V_{\eff}^{\gh+\mathrm{t}}$ and $\thetaH^{\min}=\pi$ into
(\ref{higgs-mass}), we obtain $m_h \simeq 119.7
\mathrm{\GeV}$.
This value is slightly larger than the lower-bound $m_h \ge
114\text{\GeV}$ from the LEP experiment \cite{Barate:2003sz}.

\subsection{$SU(3)$ model with an adjoint fermion}
\par
\begin{figure}[htbp]
\centerline{\includegraphics[width=0.4\linewidth]{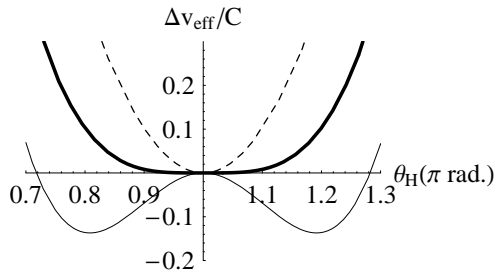}}
\caption{The plot of 
$\Delta v_{\eff}(c_\ad,\thetaH)/{\cal C}$ (see text), for $c_\ad = 0.50$
(dashed), $0.413$ (thick solid) and $0.35$ (thin solid).}\label{su3pt}
\end{figure}

In this subsection, we add an $SU(3)_w$ adjoint fermion $\Lambda$ into
the model to break unwanted $U(1)_{9}$ symmetry.\footnote{One of another
way to break the symmetry is to assign quarks and leptons into larger
representations of $SU(3)_w$, as discussed in \cite{Cacciapaglia}.}
When we turn off the gauge VEV in the AB-gauge, the $Z_2$ boundary
conditions $ \tilde{\Lambda}(y_i-y) = \gamma_5 \tilde{P}_i
\tilde{\Lambda}(y_i+y) \tilde{P}_i^\dag $ with $\tilde{P}_{i=0,1}$
defined in (\ref{su3bc}).
This boundary condition projects out the half of the zero mode of
$\Lambda^{(1-8)}$, and leaves the four left- and four right-handed
massless fermions.
Once we turn on the gauge VEV in the direction of $\lambda_7$, pairs of
left-handed and right-handed fermions
$(\Lambda^{(1)}_R,\Lambda^{(4)}_L)$,
$(\Lambda^{(2)}_R,\Lambda^{(5)}_L)$,
$(\Lambda^{(6)}_L,\Lambda^{(9)}_R)$, and pairs with opposite chiralities
(e.g. $(\Lambda^{(1)}_L,\Lambda^{(4)}_R)$) yields massive fermions:
$\Lambda^{(1\leftrightarrow4)}$, $\Lambda^{(2\leftrightarrow5)}$ and
$\Lambda^{(6\leftrightarrow9)}$, respectively.
The $n$-th KK mass the field $\Phi=\Lambda^{(i)} $ is
given by $m_{\Phi,n} = (k/z_1) x_n(\alpha_\Phi,n_\Phi \thetaH)$,
where $\alpha_{\Phi}$ and $n_{\Phi}$ are
\begin{eqnarray}
\begin{tabular}{lcc}
\hline\hline
$\Phi$ & $\alpha_\Phi$ & $n_{\Phi}$ 
\\
\hline
$\Lambda^{(1\leftrightarrow4,2\leftrightarrow5)}$ 
& $\frac{1}{2}+c_{\ad}$ & 1
\\
$\Lambda^{(9\leftrightarrow6)}$ 
& $\frac{1}{2}+c_{\ad}$ & 2
\\
$\Lambda^{(7,10)}$ 
& $\frac{1}{2} + c_{\ad}$ & 0
\\
\hline
\hline
\end{tabular},
\end{eqnarray}
with $c_{\ad}$ being the bulk mass parameter of the adjoint fermion $\Lambda$.
With these values we obtain the contribution to the effective
potential from the adjoint fermion, which is given by
\begin{eqnarray}
V_{\mathrm{adjoint}}(c_{\ad},\thetaH)
&=&
 -4 \C \cdot 
 \left[
 v(2\thetaH, \tfrac{1}{2} + c_{\ad}) + 2v(\thetaH, \tfrac{1}{2} + c_{\ad})
 \right].
\end{eqnarray}
Hence the total effective potential is given by
\begin{eqnarray}
V^{\mathrm{gh+t+ad}}_{\eff}(c_{\ad},\thetaH) 
\equiv
V_{\gh}(\thetaH) + V_{\mathrm{top}}(\thetaH) +
V_{\mathrm{adjoint}}(c_\ad,\thetaH)).
\label{total-Veff}
\end{eqnarray}
In FIG. \ref{su3pt}, shapes of 
$\Delta v_{\eff}(c_\ad,\thetaH) \equiv \left[
 V^{\gh+\mathrm{t}+\ad}_{\eff}(c_\ad,\thetaH) - 
 V^{\gh+\mathrm{t}+\ad}_{\eff}(c_\ad,\pi)\right]$
for various $c_{\ad}$ are shown.
%
%
When $|c_{\ad}| \lesssim 0.413$, the effective potential has a minimum
at $\thetaH = \thetaH^{\min}$ where $0 < \thetaH^{\min} < \pi$.
The Wilson-line phase $\langle W \rangle =
\exp(i\thetaH^{\min}\lambda_7)$ cannot commute with $T^9$ if $0 <
\thetaH^{\min} < \pi$.
Therefore the $U(1)_{9}$ gauge symmetry is broken and we obtain the
breaking : $SU(2)_w \times U(1)_8 \to U(1)_{10}$.
When the bulk mass of the adjoint fermion becomes large: $|c_{\ad}| \gtrsim
0.413$, the contribution from adjoint fermion $V_{\rm adjoint}$ becomes
negligible.
Thus the effective potential has the global minimum at $\thetaH = \pi$
and the $U(1)_{9}$ remains unbroken, as we have seen in the previous
subsection.
%
\par
When $U(1)_{9}$ is broken, the massive ``$Z$-boson'' should be
identified with lowest KK mode of $A_\mu^{(6\leftrightarrow9)}$. The
$Z$-boson mass $m_Z$ is given by
\begin{eqnarray}
m_Z =
 \mu_W(1,\thetaH,2)
 \simeq \mubar_W(1,\thetaH,2) =  m_W |\sin (\thetaH) \csc(\thetaH/2) |.
\label{zmass-eq}
\end{eqnarray}
The ratio $m_Z/m_W$ depends on the Wilson-line phase $\thetaH$ and, is a
monotonically-decreasing function of $\thetaH$ for $0 \le \thetaH \le
\pi$, and vanishes when $\thetaH = \pi$.
The massless gauge boson which appears when $\thetaH = \pi$ is the gauge
boson $A_\mu^{(9)}$ of $U(1)_9$ symmetry.
At the limit $\thetaH \to 0$, $m_Z$ and $m_W$ satisfy a relation $m_Z =
2 m_W$. This relation can be seen in the case of flat extra
dimension \cite{Manton,Murayama,Wudka06}.
In the case of the warped extra dimension with $kR > 0$, $m_Z$ ($m_W$)
dependence on $\thetaH$ is $m_Z \propto \sin \thetaH$ ($m_W \propto
\sin\thetaH/2$), whereas in flat dimension $m_Z \propto 2\thetaH$ and
$m_W \propto \thetaH$.
When $k \to 0$, the small correction to the approximation in
Eq.~(\ref{zmass-eq}) becomes no longer negligible. Thus we can expect
that in the small $k$ limit the ratio $m_Z/m_W$ approaches to one in the
case of flat extra dimension.
We should note that the change of $m_Z/m_W$ occurs at tree level by
varying $\thetaH$, unlike the radiative correction to the $T$-parameter.
%
\par
Now we estimate the lowest KK-masses of fermions $\Lambda^{(i
 \leftrightarrow j)}$, $(i,j)=(1,4),(2,5),(6,9)$:
 $m_{\ad}^{i\leftrightarrow j}$.
Recalling Eqs. (\ref{fieldmass}) and (\ref{Wmass}), we obtain 
$
m_{\ad}^{(1\leftrightarrow4)}
= m_{\ad}^{(2\leftrightarrow5)} 
= \mu_W(\frac{1}{2}+ c_{\ad}, \thetaH,1) \equiv m_{\ad1} 
$
and
$
 m_{\ad}^{(6\leftrightarrow9)} = 
\mu_W(\frac{1}{2} + c_{\ad},\thetaH, 2)  \equiv m_{\ad2}
$.
We should note that a relation $m_{\ad1}/m_{\ad2} \simeq m_Z/m_W$ holds
as long as all $m_{\ad1,\ad2,Z,W}$ are sufficiently smaller than $m_{KK}$.
Two massless fermions with opposite chirality: $\Lambda_{L,0}^{(7)}$ and
$\Lambda_{R,0}^{(10)}$ remain at tree level.
%
%
As an another physical quantity we can obtain the $n$-th Kaluza-Klein
photon $m_{\gamma,n}$ mass which is given by
$ m_{\gamma,n} = (k/z_1) x^{(0)}_n$,
where $x^{(\alpha)}_n$ is the $n$-th smallest positive solution of
$F_{\alpha,\alpha}(1/z_1,x)=0$ and $x_1^{(0)} \simeq 2.4466$.
%
\par
\begin{figure}[tbp]
\caption{Plots of 
$m_{\ad1}$ (thin dashed at the upper-right corner),
$m_{\ad2}$ (thick dashed),
$m_Z$ (thin solid) and
$m_h$ (thick solid) for $c_{\ad}$ : $0.45 \le |c_{\ad}| \le 0.55$,
with $kR=12.0$ and $\alpha_W=0.032$.}\label{fig-masslargecad}
\begin{center}
\includegraphics[width=0.5\linewidth]{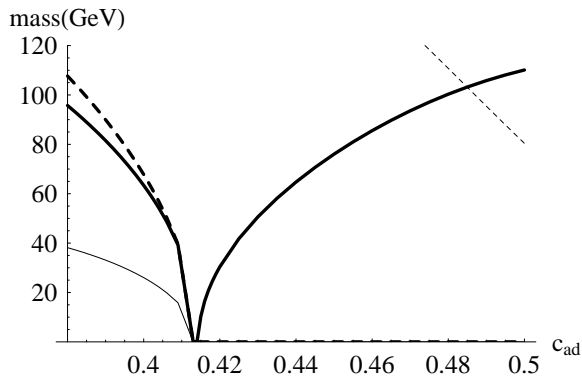}
\end{center}
\end{figure}

In this model, by fixing values of unknown parameters $kR$ and
$c_{\ad}$, the global minimum of the effective potential is determined
and we can obtain the value of $\thetaH^{\min}$ at which the effective
potential has the global minimum.
From $c_{\ad}$ and $\thetaH^{\min}$ (and effective potential
(\ref{total-Veff})), we can calculate 1-loop Higgs mass $m_h$ and tree
level masses of vector bosons $m_W$, $m_Z$ and of adjoint fermions
$m_{\ad1,\ad2}$, 1st KK-photon mass $m_{\gamma,1}$, and the KK scale $m_{KK}$.
In TABLE.~\ref{masstable}, we have summarized these masses for specific
values of $|c_{\ad}| \le 0.4$.
\begin{table}[tbp]
\caption{ The Wilson-line phase $\thetaH^{\min}$ at which the effective
potential has the global minimum, the mass of Higgs at 1-loop order, tree
level masses of $Z$-boson, adjoint fermions $m_{\ad1}$,$m_{\ad2}$, KK
scale $m_{KK}$, and 1st Kaluza-Klein mass of the photon $m_{\gamma,1}$, for
various bulk mass $c_{\ad}$ of the adjoint fermion, with $kR=12.0$,
$\alpha_W=0.0320$ and $m_W=80.4$ \GeV{} as given parameters.  }
\label{masstable}
\begin{ruledtabular}
\begin{tabular}{ccccccccc}
$|c_{\ad}|$ & $0.0$ & $0.1$ & $0.2$ & $0.3$ & $0.35$ & $0.4$ 
\\
\hline  
$\thetaH^{\min}$[$\pi$ rad.]
   & $0.733$ & $0.736$ & $0.747$ & $0.775$ &$0.808$ & $0.897$ 
\\
$m_h$[\GeV] & $233$ & $226$ & $204$ & $163$ & $127$ & $63.4$ 
\\
$m_Z$[\GeV] & $65.4$ & $64.7$ & $62.2$ & $55.6$ & $47.6$ & $26.0$ 
\\
$m_{\ad1}$[\GeV] & $436$ & $421$ & $280$ & $316$ & $274$ & $223$ 
\\
$m_{\ad2}$[\GeV] & $318$ & $305$ & $268$ & $201$ & $151$ & $67.6$ 
\\
$m_{\gamma,1}$[\GeV] & $935$ &$932$ & $926$ & $910$ & $894$ & $865$
\\
$m_{KK}$[\GeV] & $1201$ &$1198$ & $1189$ & $1169$ & $1148$ & $1111$
\end{tabular}
\end{ruledtabular}
\end{table}
%
%
In FIG.~\ref{fig-masslargecad}, we have shown the masses
$m_{\ad1}$, $m_{\ad2}$, $m_Z$, and $m_h$ for $|c_{\ad}| \ge 0.38$
with $kR=12$, $\alpha_W = 0.032$ and $m_W=80.4$ \GeV.
In the region $|c_{\ad}| \lesssim 0.413$, all of these masses are
monotonically decreasing function of $|c_{\ad}|$. 
When $|c_{\ad}| \gtrsim 0.413$, we obtain $\thetaH^{\min} = \pi$. Thus
$m_{\ad2}$ and $m_Z$ vanish in this region.
The mass of the Higgs also decreases with increasing $|c|$ for larger
$|c_{\ad}|$ as long as $|c_{\ad}| \lesssim 0.413$.
For $|c_{\ad}| \gtrsim 0.413$, however, the mass of Higgs increases and
closes to $\sim 120\GeV$, because the contribution from the top quark
becomes dominant in this region.

So that $m_W = 80.4$ \GeV{} and $m_Z=91.2$ \GeV{} satisfy the relation
(\ref{zmass-eq}), $\thetaH^{\min} \simeq 0.616\pi$ is required.
This value of $\theta_H^{\min}$ is, however, smaller than the lower
bound $\thetaH^{\min} \gtrsim 0.733\pi$ which is obtained by solving
Wilson-line dynamics.
Furthermore, it seems unlikely to push up (down) the $m_Z$
($\thetaH^{\min}$) just by introducing more adjoint fermions, because
the minimum of the $V_{\rm adjoint}(\thetaH)$ locates at $\thetaH \sim
0.689\pi$ and we cannot make $\thetaH^{\min}$ smaller than $0.689\pi$
only by introducing more adjoint fermion into the model.
In the following subsections we try to push up (down) $m_Z$
($\thetaH^{\min}$), by adding some scalar fields or twisted fermions.

\subsection{Adding Scalar Fields}\label{adding_scalar}
The action of a bulk scalar field $S$ can be
written as \cite{GP}
\begin{eqnarray}
\L_s &=& \int\!\! d^4x \!\int_{-\pi R}^{\pi R} \!\! dy\, \sqrt{|G|} 
\left\{
(D_M S)^\dag (D^M S) - (s_S k^2 +t_S \sigma'') S^\dag S
\right\},
\end{eqnarray}
where $s_{S}$ and $t_{S}$ are the bulk and boundary mass of the scalar
field $S$, respectively.\footnote{For simplicity, we do not include
scalar quartic terms. Therefore, we do not consider the case in which
the scalar fields develop VEVs and cause Higgs mechanism here. The cases
in which the Higgs mechanism and the dynamical gauge-Higgs unification
coexist are discussed in \cite{Sakamoto_gauge-higgs}.}
When $s_S$ and $t_S$ satisfy the relation:
\begin{eqnarray}
2-t_{S} = \sqrt{4+s_{S}} = \nu,
\label{scalar-mass-condition}
\end{eqnarray}
the KK-mass spectrum of such scalar becomes identical to a fermion field
with the bulk mass $c=\pm(\nu - 1/2)$, and the contribution to the
effective potential par degrees of freedom turns out to be just same
magnitude but with opposite sign as the one from the fermion in the same
representation of $SU(3)_w$.
Hereafter we consider the case where boundary- and
bulk-masses of scalar fields satisfy the condition
(\ref{scalar-mass-condition}).
%
\par
\begin{figure}
\caption{The dependence of $\thetaH^{\min}$ as the minimum of
$V_{\eff}^{{\rm gh}+{\rm t}+\ad+{\rm s}}$ on the scalar contribution
$\xi$, for $c_{\ad}=0.0$ (thick solid), $0.3$ (thin solid), $0.4$ (thin
dashed) and $0.45$ (thick dashed).
FIG.\ref{xi-depend}-(b) is a close-up view of \ref{xi-depend}-(a).  In
both (a) and (b), the solid horizontal line shows $\thetaH = 0.616\pi$.
}\label{xi-depend}
\includegraphics[width=0.4\linewidth]{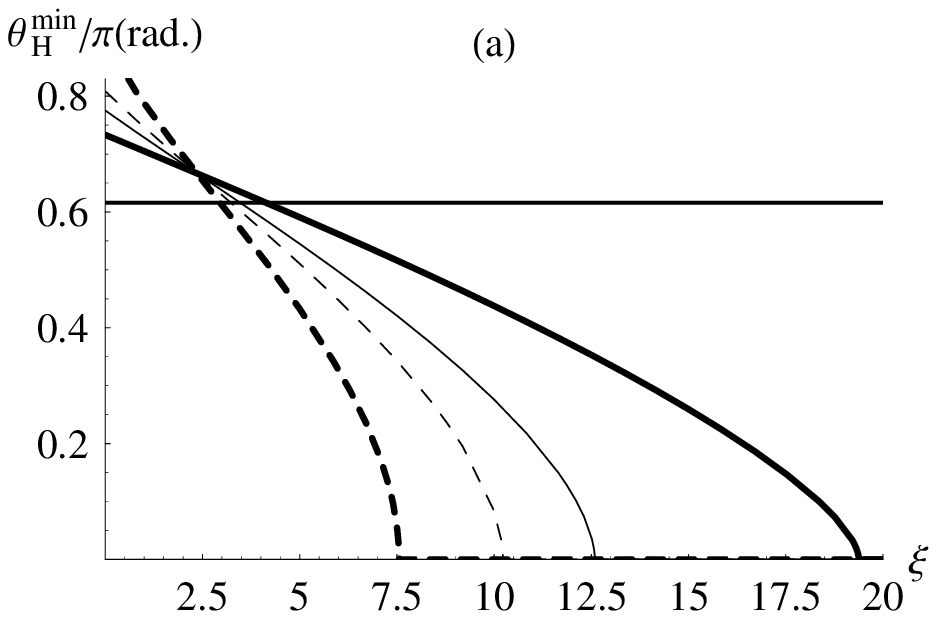}
\includegraphics[width=0.4\linewidth]{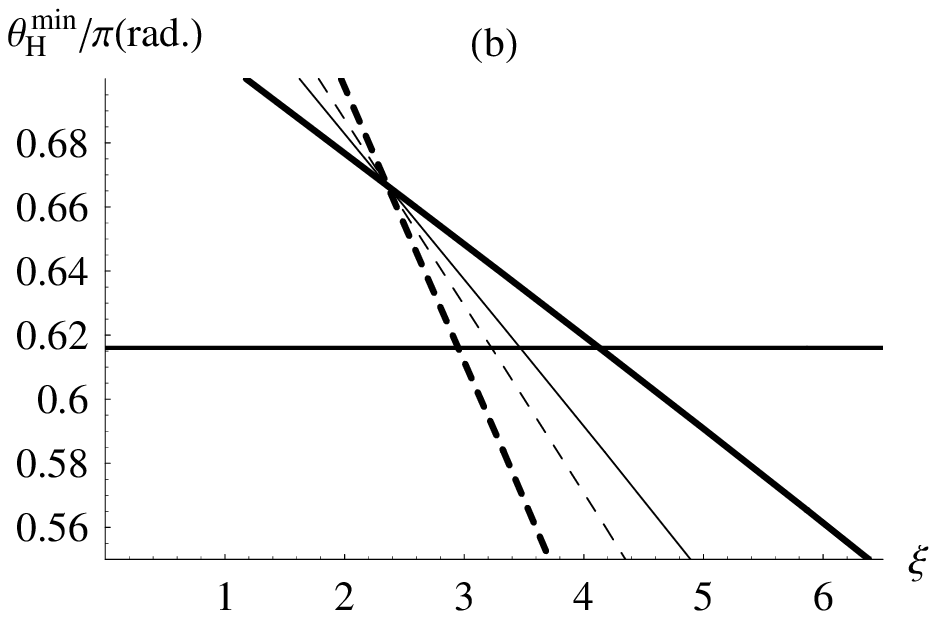}
\end{figure}

Now we propose a way to lift up (put down) the $Z$-boson mass
($\thetaH^{\min}$) by introducing scalar fields into the model.
We add one or more scalar fields $S^a_{f}$($f=1,\dots,N_s$, $a=1,2,3$),
which are in the fundamental representation of $SU(3)_w$, into the model.
The contribution to the effective potential from such scalar fields is
given as
\begin{eqnarray}
V_{\rm scalar} = +2\C \sum_{f=1}^{N_s} v(\tfrac{1}{2} + b_f,\thetaH),
\label{Veff-scalar}
\end{eqnarray}
where $b_f$ ($f=1,\dots,N_s$) are mass parameters of the scalar $S_f^a$
and related with the boundary and bulk mass parameters $t_{S_f}$,
$s_{S_f}$ by  $t_{S_f} = b_f \mp 3/2$ and $s_{S_f} = b_f^2 \pm
b_f - 15/4$.
The total effective potential is given by
$
V_{\eff}^{{\rm gh}+{\rm t}+\ad+{\rm s}} \equiv 
 V_{\eff}^{{\rm gh}+{\rm t}+\ad} + V_{\rm scalar}$.
For simplicity, we use an approximation of Eq.~(\ref{Veff-scalar}),
which is given by
\begin{eqnarray}
V_{\rm scalar} \simeq - \xi \cdot \C \cdot \Re\,\Li_5(e^{i\theta}),
\label{scalar-approx}
\end{eqnarray}
where we have introduced a dimensionless parameter $\xi=\xi(b_f)$, $0 \le \xi
\lesssim \frac{3}{2} N_s$.
Since $V_{\rm scalar}(\thetaH)$ has the global minimum at
$\thetaH = 0$, we can shift the location of the minimum
$\thetaH^{\min}$ of $V_{\eff}^{{\rm gh}+{\rm t}+\ad+{\rm s}}$ to $ 0 $
by increasing the value of $\xi$.
Thus in this model we can change the ratio $m_Z/m_W$ by tuning adjoint
fermion mass $c_{\ad}$ and the scalar field contribution parameterized by
$\xi$.
In FIG.~\ref{xi-depend}(a) and (b), we have shown how $\thetaH^{\min}$
depends on $\xi$.
FIG.~\ref{xi-depend} tells that $\thetaH^{\min} \simeq 0.616\pi$ can be
achieved by tuning $\xi$ for any value of $c_{\ad}$.
Thus for a given value of $c_{\ad}$, we can define $\xi^c =
\xi^c(c_{\ad})$ such that $V_{\eff}^{{\rm gh}+{\rm t}+\ad+{\rm s}}$ has
the global minimum at $\thetaH = 0.616\pi$ with $\xi = \xi^c$.
In TABLE.~\ref{tbl:xi-depend_masses}, we summarized the value of $\xi^c$
for each value of $c_{\ad}$. And we re-calculate masses $m_h$,
$m_{\ad1,2}$ with $\xi^c$ and $\thetaH^{\min} = 0.616\pi$.
\begin{table}
\caption{Mass spectrum of $SU(3)_w$ models with an adjoint fermion and
 scalar fields which makes $\thetaH=\thetaH^c\equiv0.616\pi$.}\label{tbl:xi-depend_masses}
\begin{ruledtabular}
\begin{tabular}{ccccccc}
$|c_{\ad}|$ & $0.0$ & $0.1$ & $0.2$ & $0.3$ & $0.35$ & $0.4$ \\
\hline
$\xi^c$     & $4.13$& $4.06$& $3.84$& $3.47$& $3.23$& $2.95$ \\
\hline
$m_h(c_{\ad},\thetaH^c,\xi^c)$[\GeV] &
$291$ & $285$ & $266$ & $231$ & $205$ & $169$ \\
$m_{\ad1}(c_{\ad},\thetaH^c)$ [\GeV]& 
$349$ & $342$ & $320$ & $279$ & $249$ & $209$ \\
$m_{\ad2}(c_{\ad},\thetaH^c)$ [\GeV]& 
$396$ & $388$ & $363$ & $317$ & $283$ & $238$  
\end{tabular}
\end{ruledtabular}
\end{table}
The KK-scale $m_{KK}$ and the first KK-photon mass at
$\thetaH=0.616\pi$ are $1331$\GeV{} and $1037$\GeV, respectively.
\par
One may wonder why the Higgs mass becomes larger nevertheless scalar
field may cancel the fermion's contribution.
The reason can be explained as follows.
First, we remember that $V_{\rm adjoint}(\thetaH)$ and $V_{\rm
scalar}(\thetaH)$ have similar shape to $\cos2\thetaH$ and
$-\cos\thetaH$, respectively.
Then the contribution of the adjoint fermion to the Higgs mass (i.e. the
curvature of $V_{\rm adjoint}(\thetaH)$) has the maximum around at
$\thetaH \sim \pi/2$, whereas one of fundamental scalars vanishes at
$\thetaH \sim \pi/2$.
Thus the Higgs mass becomes small when the $\thetaH$ closes to $\pi/2$.

\subsection{Twisted Fermion}
An alternative way to shift the Z-boson mass in the $SU(3)_w$ model is
to introduce one or more ``twisted fermions'' $\psi_{\rm tw}$, which
boundary condition is twisted even in AB-gauge.
As an example, we consider the case where $\psi_{\rm tw}$ is in the
fundamental representation of $SU(3)_w$. As a possible boundary
condition of $\psi_{\rm tw}$ in the AB-gauge, we define
\begin{eqnarray}
\tilde\psi_{\rm
tw}(x,y_i+y) = 
\eta_{\rm tw} \gamma_5
\tilde{P}_{tw,i} 
 \tilde\psi(x,y_i-y) \quad (i=0,1),
\\
\tilde{P}_{tw,0} = \tilde{P}_{0},
\quad
\tilde{P}_{tw,1}= \exp(i\varphi\lambda_7)\tilde{P}_{1}\exp(-i\varphi\lambda_7),
\end{eqnarray}
where $\tilde{P}_{i}$ is defined in (\ref{su3bc}) and $\eta_{\rm tw} = \pm1$.
When we introduce $N_t$ copies of such twisted-fermions $\psi_{\rm tw}^i$
($i=1,...,N_t$) with $\varphi = \pi$ into the model, the contribution to
the one-loop effective potential induced from such fermions is given by
\begin{eqnarray}
V_{\rm tw} = -4\C \sum_{i=1}^{N_t} 
 v(\tfrac{1}{2}+c_{\rm tw}^{i},\thetaH+\pi),
\end{eqnarray}
where $c_{\rm tw}^{i}$ is the bulk mass parameter of $\psi_{\rm tw}^i$.
Since $v(\nu,\theta+\pi) \approx - v(\nu,\theta)$, we can make use of
the result of Sec.~\ref{adding_scalar} by replacing $V_{\rm scalar}$ in
(\ref{scalar-approx}) with $V_{\rm tw}$ and 
$N_s/2$ with $N_t$.
%
%
\section{Summary and Comments}\label{sec-summary}
%
In the present paper, we investigated the dynamical gauge-Higgs
unification in the RS space-time.
%
%
In Sec. \ref{sec-hosotanimech} we consider the SU(2) gauge theory in the RS
space-time.
We calculate one-loop effective potential with respect to the
Wilson-line phase.
Especially we clarified the contribution from a fermion with the bulk
mass parameter $c$.
The obtained effective potential properly inter/extrapolates the results
which are known for $|c|=0,\,1/2$.
We see that the gauge symmetry can be broken by dynamically-induced
Wilson line when we introduce an adjoint fermion into the model and that
the breaking pattern of the gauge symmetry depends on the bulk mass of
the adjoint fermion.
%
\par
In Sec.\ref{sec-su3model} we consider $SU(3)_w$ gauge models in the RS as
toy models of 5D extensions of the electroweak theory.
We found that it is possible to break $SU(2)\times U(1)$ to $U(1)$ by
introducing an adjoint fermion.
The large mass hierarchy among quarks and leptons are naturally obtained
by adjusting their bulk mass parameters of the order of unity.
We calculate one-loop Higgs mass numerically. In the $SU(3)_w$ model the
Higgs mass can be changed from zero to $\sim 290\,\GeV$ with $kR=12.0$.
We have also estimated mass spectrum of this model for various value of
mass of the adjoint fermion, which are determined by the RS parameter $kR$
and bulk mass parameter of the adjoint fermion.
Interestingly, these predicted masses of Higgs and new fermions are
in the range where LHC experiment can explore.
%
%
In this model, we see that the ratio of $W$-boson mass to the $Z$-boson
mass varies with respect to the Wilson-line phase. This occurs at tree
level.
We find the way to tune the ratio $m_Z/m_W$ to satisfy the
realistic one $\sim 91.2/80.4$, by introducing $SU(3)_w$
fundamental scalar fields with the bulk and boundary mass terms which
satisfy the relation (\ref{scalar-mass-condition}), or fermions with
twisted boundary condition.
Unfortunately, the $SU(3)_w$ model still has some problems, e.g., varying
$m_Z/m_W$ may occur due to the lack of custodial symmetry in this model,
some SM and non-SM fermions remain massless, and quarks and leptons
cannot have correct isospin and hypercharge simultaneously.
To obtain a more realistic model of the EWGHU in the RS space-time, the
choice of the enhanced electroweak symmetry and the assignment of matter
contents should be reconsidered.
%
\par
We have seen that in the $SU(3)_w$ model a quark with small
bulk masses has a heavy lowest KK state\footnote{It is not
true in the case of flat extra dimension, where a fermion with large
bulk mass tends to have large lowest KK mass.}, and that such heavy
fermions have large contribution to the effective potential of Higgs
because of the smallness of bulk masses.
The converse is also true; fermions with large bulk mass term have light
lowest KK masses and have small contributions to the effective potential.
It remind us the qualitative similarity with the CW mechanism; In CW
mechanism the contribution to the Higgs potential from a heavy quark
loop is large because a heavy fermion has large Yukawa coupling to the
Higgs field.
However, we have to note that in the dynamical gauge-Higgs unification
the quadratic divergence of Higgs mass is absent, thanks to the gauge
symmetry in the higher-dimensional space-time.
%
\begin{acknowledgments}
\noindent 
Author will thank Darwin Chang for fruitful discussion.
This work was initiated during author's stay at National Tsing-Hua University.
Author also thanks
National Center for Theoretical Science in Hsinchu for warm hospitality.
This work is partly supported by National Science Council of
R.O.C. under Grant No. 93-2112-M-007-015.
\end{acknowledgments}
\appendix
\section{Approximation Formulas}
\label{sec-approx}
When the 4D (or lowest KK mode) mass of a field $\Phi$ is sufficiently
smaller than $m_{KK}$, i.e., $\pi^2 m_{\Phi,0}^2 \ll m_{KK}^2$, we can
use an approximation formula for $m_{\Phi,0}$, as discussed in
\cite{Hosotani06}.  The mass $m_{\Phi,0}$ is approximated by
\begin{eqnarray}
m_{\Phi,0} = \mubar(\alpha_\Phi,\thetaH,n_\Phi) 
\left\{1 + {\cal O} \left(\frac{\mubar^2 \pi^2}{m_{KK}^2} 
\right)\right\},
\end{eqnarray}
where $\mubar(\alpha,\theta,n)$ is defined by
\begin{eqnarray}
\mubar(\alpha,\thetaH, n) &\equiv& k 
\sqrt{
\frac{\alpha(\alpha-1)}{z_1\sinh [\alpha k\pi R] \sinh [(\alpha-1)k \pi R]}
} \cdot \left|\sin\frac{n\thetaH}{2}\right|.
\label{fieldmass}
\end{eqnarray}
The mass of the weak boson $m_W$ is give by $m_W = \mu(1,\thetaH,1)$,
and can be approximated by\footnote{Here we have taken the limit:
$\lim_{\alpha \to 1} \mubar(1,\thetaH,1)$.}
\begin{eqnarray}
m_W 
=
\frac{m_{KK}}{\pi}
\sqrt{\frac{2}{k\pi R}}\cdot
\left| \sin \frac{\thetaH}{2} \right|
\left\{1 + {\cal O} \left(\frac{m_W^2 \pi^2}{m_{KK}^2}\right)\right\}
.\label{Wmass}
\end{eqnarray}
From Eqs. (\ref{fieldmass}),(\ref{Wmass}),
the lowest KK mass $m_{\Phi,0}$ is also approximated (in terms of $m_W$)
by
\begin{eqnarray}
m_{\Phi,0}
 = \mubar_W(\alpha_{\Phi},\thetaH,n_{\Phi}) 
\left\{1 + {\cal O} (m_{\Phi,0}^2 \pi^2/m_{KK}^2)\right\},
\end{eqnarray}
where $\mubar_W$ is defined by
\begin{eqnarray}
\mubar_W(\alpha,\thetaH, n)
&\equiv&
m_W
\frac{\mubar(\alpha,\thetaH,n)}{\mubar(1,\theta,1)}
\nonumber\\
&=&
m_W
\sqrt{
\frac{z_1 \alpha(\alpha-1)kR\pi}{
2 \sinh[\alpha kR\pi]\sinh[(\alpha-1)kR\pi]}
}
\cdot
\left|\sin\frac{n\thetaH}{2} \csc\frac{\thetaH}{2}\right|.
\end{eqnarray}
Here we should note that 
$\mubar_W(\alpha,\thetaH,1)$ is independent of the $\thetaH$.
\par
As for $m_{KK}$, by using $W$-mass formula (\ref{Wmass}) inversely we
obtain
\begin{eqnarray}
m_{KK} \simeq
 \pi m_W \sqrt{\pi k R/2} \left|\csc(\thetaH/2) \right|.
\label{KKmass}
\end{eqnarray}
\par
\begin{figure}
\caption{$\mu_W(\frac{1}{2}-c,\thetaH,1)$  and
 $\mubar_W(\frac{1}{2}-c,\thetaH,1)$ with
 fixed $m_W=80.4\GeV$ and $kR=12$ and various $c$ are plotted.
From the top, the thick solid, thick dashed, thin dashed and thin solid
curve [horizontal line] show $\mu_W(\frac{1}{2}-c,\thetaH,1)$
[$\mubar_W(\frac{1}{2}-c,\thetaH,1)$] for $c=0.0,0.3,0.4$ and $0.45$,
respectively.
The lowest single horizontal line indicates $m_W = 80.4\GeV$.}\label{mubarmu}
\includegraphics[width=4in]{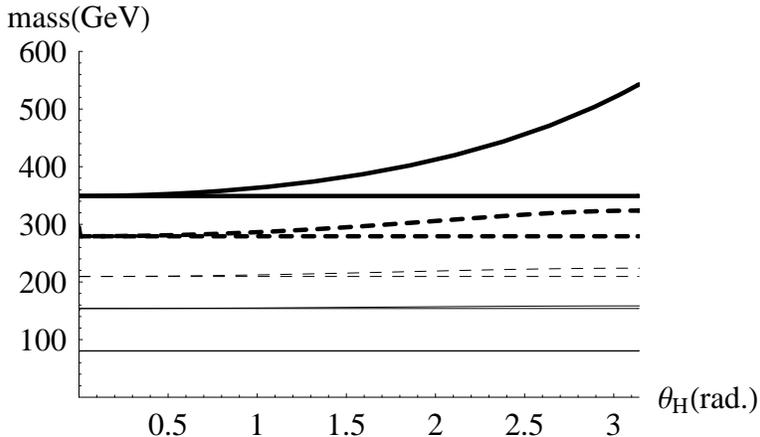}
\end{figure}

We should also note that Eq. (\ref{fieldmass}) is an approximation of
$\mu(\alpha,\thetaH,n)$ and valid valid only when $\mubar \ll m_{KK}$.
When $\alpha$ close to $1/2$, the difference between
$\mu(\alpha,\thetaH,n)$ and $\mubar(\alpha,\thetaH,n)$ becomes large (see
FIG. \ref{mubarmu}).


%
%
\end{document}